\documentclass[11pt]{iopart}
\newcommand{\email}[1]{\ead{#1}}
\newcommand{\homepage}[2]{}
\newcommand{\tfrac}[2]{\frac{#1}{#2}}
\newcommand{\affiliation}[1]{\address{#1}}
\newcommand{\eqref}[1]{(\ref{#1})}
\newcommand{\binom}[2]{\left( \begin{array}{c}{#1}\\{#2}\end{array} \right)}

\newcommand{\openone}{\mathbbm1}

\usepackage{graphics}
\usepackage{epsfig}
\usepackage{bbm}
\usepackage[latin1]{inputenc} 
\usepackage{subfigure}
\usepackage{subeqn} 

\newcommand{\EQ}[1]{\eqref{#1}}
\newcommand{\EQS}[1]{\eqref{#1}}
\newcommand{\EQLL}[1]{equation~\eqref{#1}}
\newcommand{\EQL}[1]{Equation~\eqref{#1}}
\newcommand{\SEC}[1]{section~\ref{#1}}

\newcommand{\FIG}[1]{figure~\ref{#1}}
\newcommand{\CFIG}[1]{Figure~\ref{#1}}
\newcommand{\REF}[1]{reference~\cite{#1}}
\newcommand{\REFS}[1]{references~\cite{#1}}
\newcommand{\REFL}[1]{Reference~\cite{#1}}
\newcommand{\REFSL}[1]{References~\cite{#1}}

\newcommand{\be}{\begin{equation}}
\newcommand{\ee}{\end{equation}}
\newcommand{\eea}{\end{eqnarray}}
\newcommand{\bea}{\begin{eqn array}}

\newcommand{\va}[1]{\ensuremath{(\Delta#1)^2}}

\newcommand{\ex}[1]{\ensuremath{\left\langle{#1}\right\rangle}}
\newcommand{\exs}[1]{\ensuremath{\langle{#1}\rangle}}

\newcommand{\eins}{\openone}

\newcommand{\ketbra}[1]{\ensuremath{| #1 \rangle \langle #1 |}}

\newcommand{\ket}[1]{\ensuremath{|#1\rangle}}

\newcommand{\bra}[1]{\ensuremath{\langle#1|}}

\newcommand{\kommentar}[1]{}
\newcommand{\trace}{{\rm Tr}}

\newcommand{\singlet}{\Psi^-}
\def\be{\begin{equation}}
\def\ee{\end{equation}}
\def\eea{\end{eqnarray}}
\def\bea{\begin{eqnarray}}

\newcommand{\phtheta}{{\theta}}

\def\@fnsymbol#1{\ifcase#1\or 1 \or 2 \or 3 \or
   \|\or \P\or ^{+}\or ^{\tsty *}\or \sharp
   \or \dagger\dagger \else\@ctrerr\fi\relax}

\begin{document}
\review[Quantum metrology from a quantum information science perspective]{Quantum metrology from a quantum information science perspective}
\date{\today}
\author{G\'eza T\'oth$^{1,2,3},$ Iagoba Apellaniz$^1$}
\affiliation{$^1$ Department of Theoretical Physics, University of the Basque Country  UPV/EHU,
P.O. Box 644, E-48080 Bilbao, Spain}
\affiliation{$^2$ IKERBASQUE, Basque Foundation for Science, E-48011 Bilbao, Spain}
\affiliation{$^3$ Wigner Research Centre for Physics, Hungarian
Academy of Sciences, P.O. Box 49, H-1525 Budapest, Hungary}
\email{toth@alumni.nd.edu}
\affiliation{URL: \tt{http://www.gtoth.eu}}
\email{iagoba.apellaniz@gmail.com}
\begin{abstract}
We summarise important recent advances in quantum metrology, in connection to experiments in cold gases, trapped cold atoms and photons.  First we review simple metrological setups, such as quantum metrology with spin squeezed states, with Greenberger-Horne-Zeilinger states, Dicke states and singlet states. We calculate the highest precision achievable in these schemes. Then, we present the fundamental notions of quantum metrology, such as shot-noise scaling, Heisenberg scaling, the quantum Fisher information and the Cram\'er-Rao bound.
Using these, we demonstrate that entanglement is needed to surpass the shot-noise scaling in very general metrological tasks with a linear interferometer.
We discuss some applications of the quantum Fisher information,
such as how it can be used  to obtain a criterion for a quantum state to be a macroscopic 
superposition.
We show how it is related to the 
the speed of a quantum evolution, and how it appears in the theory of the quantum Zeno effect.
Finally, we explain how uncorrelated noise limits the highest achievable precision in very general metrological tasks.
\end{abstract}
This article is part of a special issue of {\it Journal of Physics A: Mathematical
and Theoretical} devoted to `50 years of Bell's theorem'.



\maketitle

\tableofcontents

\makeatletter
\@mkboth{Quantum metrology from a quantum information science perspective}{Quantum metrology from a quantum information science perspective}
\makeatother



\section{Introduction}

Metrology plays a central role in science and engineering. In short, it is concerned with the highest
achievable precision in various parameter estimation tasks, and with finding measurement schemes that reach that precision. Originally, metrology was focusing on measurements using classical or semiclassical systems, such as mechanical systems described by classical physics or optical systems modelled by classical wave optics.
In the last decades, it has become possible to observe the dynamics of many-body quantum systems.
If such systems are used for metrology, the quantum nature of the problem plays an essential role
in the metrological setup.
Examples of the case above are phase measurements with trapped ions \cite{leibfried2004toward}, inteferometry with photons \cite{mitchell2004super,PhysRevLett.94.090502,nagata2007beating,PhysRevLett.98.223601} and magnetometry with cold atomic ensembles \cite{wildermuth2006sensing,PhysRevLett.98.200801,behbood2013real,koschorreck2011high,lucke2011twin,2014arXiv1405.6022M}.

In this paper, we review various aspects of quantum metrology with the intention to give a comprehensive picture
to scientists with a quantum information science background.
 We will present simple examples that, while can be used to 
explain the fundamental principles, have also been realised experimentally.
The basics of quantum metrology \cite{Giovannetti19112004,0953-4075-45-10-103001,doi:10.1142/S0219749909004839,giovannetti2011advances,wiseman2010quantum,caves2010quantum,2008ConPh..49..125D,2014arXiv1405.7703D,dunningham2006using,arXiv:1308.6092} 
can be perhaps best understood in the fundamental task of magnetometry with
a fully polarised atomic ensemble. It is easy to deduce the precision limits of the parameter estimation, 
as well as the methods that can improve the precision.
We will also consider phase estimation with other highly entangled states such as for example
Greenberger-Horne-Zeilinger (GHZ) states \cite{ghz-argument}.

After discussing concrete examples, we present a general framework for computing
the precision of the parameter estimation in the quantum case, based on the Cram\'er-Rao bound and the quantum Fisher 
information. In the many-particle case, most of the metrology experiments have been done in systems
with simple Hamiltonians that do not contain interaction terms.
Such Hamiltonians cannot create entanglement between the particles. 
For cold atoms, 
a typical situation is that the input state is rotated with some angle and this angle must be estimated.
We will show that quantum states with particles exhibiting quantum correlations, or
more precisely, quantum entanglement \cite{RevModPhys.81.865,guhne2009entanglement}, provide a higher precision than an ensemble of uncorrelated particles.
The most important question is how the achievable precision $\Delta \theta$ scales with the number of particles.
Very general derivations lead to, at best,
\begin{equation} \label{eq:shotnoise}
\va{\theta}\sim  \tfrac{1}{N}
\end{equation}
for nonentangled particles. \EQL{eq:shotnoise} is called the {\it shot-noise scaling},
the term originating from the shot-noise 
in electronic
circuits, which is due to the  discrete nature of the electric charge.
$\theta$ is a parameter of a very general unitary evolution that we would like to estimate.
On the other hand, quantum entanglement makes it possible to reach
\begin{equation} \label{eq:Heisenberg}
\va{\theta}\sim \tfrac{1}{N^2},
\end{equation}
which is called the {\it Heisenberg-scaling.}
Note that if the Hamiltonian of the dynamics has interaction terms then even a better scaling is possible  (see, e.g., \REFS{Luis20048,napolitano2011interaction,PhysRevLett.98.090401,braun2011heisenberg,PhysRevLett.100.220501,PhysRevA.77.053613,PhysRevA.76.053617}). 

All the above calculations have been carried out for an idealised situation.
When a uncorrelated noise is present in the system, it turns out that for large enough particle numbers
the scaling becomes a shot-noise scaling.
The possible survival of a better scaling under correlated noise, 
under particular circumstances, or depending on some interpretation
of the metrological task, is at the centre of attention currently. All these are strongly connected
to the question of whether strong multipartite entanglement can survive in
a noisy environment.

Our paper is organised as follows. In \SEC{sec:Examples}, we will discuss examples of 
metrology with large particle ensembles and show simple methods to obtain upper bounds on the achievable precision.
In \SEC{sec:Spin squeezing and entanglement}, we define multipartite entanglement and discuss how entanglement is needed for spin squeezing, which is a typical method to improve the precision of some metrological applications
in cold gases. 
We also discuss some generalised spin squeezing entanglement criteria.
In \SEC{sec:The quantum Fisher information}, 
we introduce the Cram\'er-Rao bound and the quantum Fisher information, and other fundamental notions 
of quantum metrology.
In \SEC{sec:Quantum Fisher information and entanglement}, we show that
multipartite entanglement is a prerequisite for maximal metrological precision in many very general metrological tasks. We also discuss how to define macroscopic superpositions, how the entanglement properties of the quantum state are related to the speed of the quantum mechanical processes and to the quantum Zeno effect.
We will also very briefly discuss the meaning of inter-particle entanglement in many-particle systems.
In \SEC{Sec:Metrology with noise}, we review some of the very exciting recent findings showing that uncorrelated noise can change the scaling
of the precision with the particle number under very general assumptions.

\section{Examples for simple metrological tasks with many-particle ensembles}
\label{sec:Examples}

In this section, we present some simple examples of quantum metrology, involving an
 ensemble of $N$ spin-$\frac{1}{2}$ particles in
an external magnetic field.
We demonstrate how simple ideas of 
quantum metrology can help to determine the precision of 
some basic tasks in parameter estimation.
We consider completely polarised ensembles, as well as GHZ states, 
symmetric Dicke states \cite{PhysRev.93.99,PhysRevLett.98.063604,lucke2011twin,hamley2012spin} and singlet states \cite{1367-2630-12-5-053007,PhysRevA.88.013626}.

First, let us explain the characteristics of the physical system we use for our discussion.
In a large particle ensemble, typically only collective quantities can be measured. 
For spin-$\frac{1}{2}$ particles, such collective quantities are the angular momentum components defined as
\be 
J_l:=\sum_{n=1}^N
j_{l}^{(n)} \ee 
for $l= x,y,z,$ where $j_{l}^{(n)}$ are the components of the
angular momentum of the $n^{\rm th}$ particle.
More concretely, we can measure the 
expectation values and the variance of the angular momentum component $J_{\vec n}=\sum_{l=x,y,z} n_l J_l,$
where $\vec n$ is a unit vector describing the component.

The typical Hamiltonians involve also collective observables, such as the Hamiltonian
describing the action of a magnetic field pointing in the ${\vec b}$-direction
\be 
\label{HB}
H_{B}=\gamma BJ_{\vec b},
\ee 
where $\gamma$ is the gyromagnetic ratio, $B$ is the strength of the magnetic field,  $\vec b$ is the direction of the field, 
and  $J_{\vec b}$ is the angular momentum component
parallel with the field.
Hamiltonians of the type \eqref{HB} do not contain interaction terms, thus starting from a product state
we arrive also at a product state. 
Interferometry with dynamics determined by \eqref{HB} is discussed in the context of SU(2) interferometers \cite{PhysRevA.33.4033}, as the $J_l$ are the generators of the SU(2) group. We will mostly study this type of interferometry in multiparticle systems, as it gives a good opportunity to relate the entanglement of many-particle states to their metrological performance.

The Hamiltonian \eqref{HB}, with the choice of $\hbar=1,$ generates the dynamics
\be 
U_\theta=e^{-iJ_{\vec n}\theta},
\ee 
where we defined the angle $\theta$ that depends on the evolution time $t$ 
\be 
\theta=\gamma Bt.
\ee 
A basic task in quantum metrology is to estimate the small parameter $\theta$
by measuring
the expectation value of a Hermitian operator, which we will denote by $M$ in the following. 
If the evolution time $t$ is a constant then estimating $\theta$ is equivalent to estimating the
magnetic field $B.$
The precision of the estimation can be characterised
with the error-propagation formula as
\begin{equation}\label{acc}
(\Delta\theta)^{2}=\frac{(\Delta M)^2}{\vert \partial_\theta \exs{M}\vert^2},
\end{equation}
where $\exs{M}$ is the expectation value of the operator $M,$
and its variance is given as
\begin{equation}
(\Delta M)^2=\exs{M^2}-\exs{M}^2.
\end{equation}
Thus, the precision of the estimate depends on how sensitive $\exs{M}$ 
is to the change of $\theta,$ and also on how large the variance
of $M$ is. Based on the formula
\eqref{acc}, one can see that the larger the slope $\vert \partial_\theta \exs{M}\vert,$ the higher the precision.
On the other hand, the larger the variance $(\Delta M)^2,$
the lower the precision.

\begin{figure}
\begin{center}
\includegraphics[width=11cm]{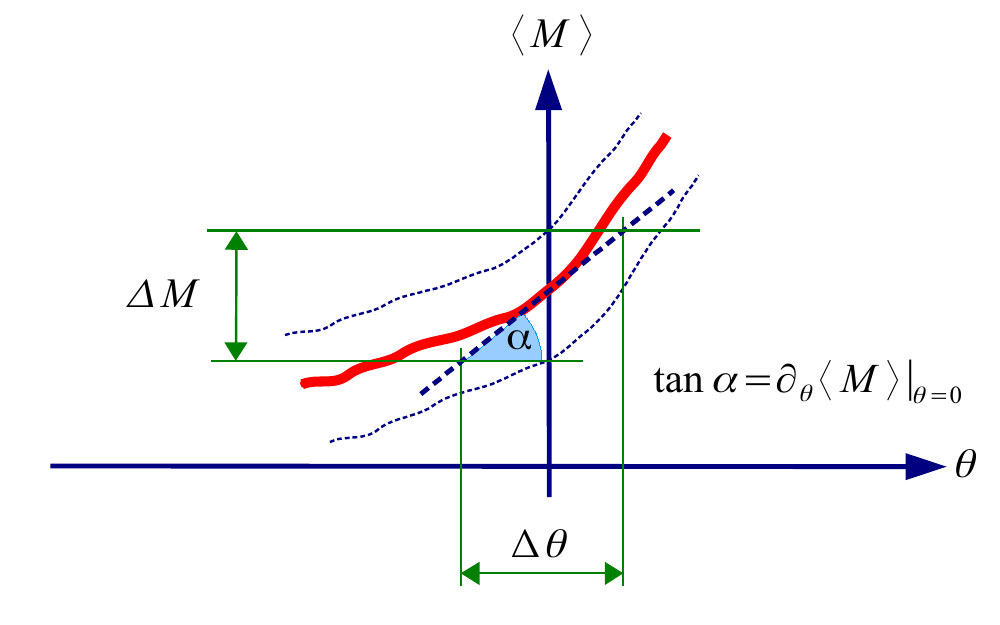} 
\caption{
Calculating the precision of estimating the small parameter $\theta$ 
based on measuring $M$ given by the error propagation formula \eqref{acc}. (solid curve) The expectation value $\ex{M}$ as a function $\theta.$ 
(dashed curves) Uncertainty of $M$ as a function of $\theta$ given as a confidence interval.
(vertical arrow) The uncertainty $\Delta M\equiv\sqrt{\va{M}}$ for $\theta=0.$ (dashed line) Tangent of the curve $\ex{M}(\theta)$ 
at $\theta=0.$ Its slope is $\partial_\theta \exs{M}.$ (horizontal arrow)  $\Delta \theta$ is the uncertainty of the parameter estimation.}
\label{fig:var}
\end{center}
\end{figure}

The formula \eqref{acc} above can be calculated for any given $\theta.$
Thus, this formalism can be used to characterise small fluctuations
around a given $\theta.$
For simplicity we will calculate the precision typically for $\theta=0.$ (To be more precise, if both the numerator and the 
denominator in \EQ{acc} are zero, then we will take the $\theta\rightarrow0$ limit instead.)
This approach is connected to the estimation theory based on the quantum Fisher information discussed in this review and could be called a {\it local} approach. 
\CFIG{fig:var} helps to interpret the quantities appearing in \EQ{acc}.
We note that the {\it global}  alternative
is the Bayesian estimation theory. 
There, the  parameter to be estimated is a random variable with a certain probability density $p(\theta).$
For a recent review discussing this approach in detail, see \REF{2014arXiv1405.7703D}. For an application, see \REF{PhysRevA.82.053804}.

Finally, note that often, instead of $(\Delta\theta)^{2}$ one calculates $(\Delta\theta)^{-2},$ which 
is large for a high precision. It scales as $\sim N$ for the shot-noise scaling, and  as $\sim N^2$
for the Heisenberg scaling. [Compare with \EQS{eq:shotnoise} and \eqref{eq:Heisenberg}.]

\subsection{Ramsey-interferometry with spin squeezed states}

\label{Sec:Spin squeezed states}
\begin{figure*}
\begin{center}
\includegraphics[width=8.6cm]{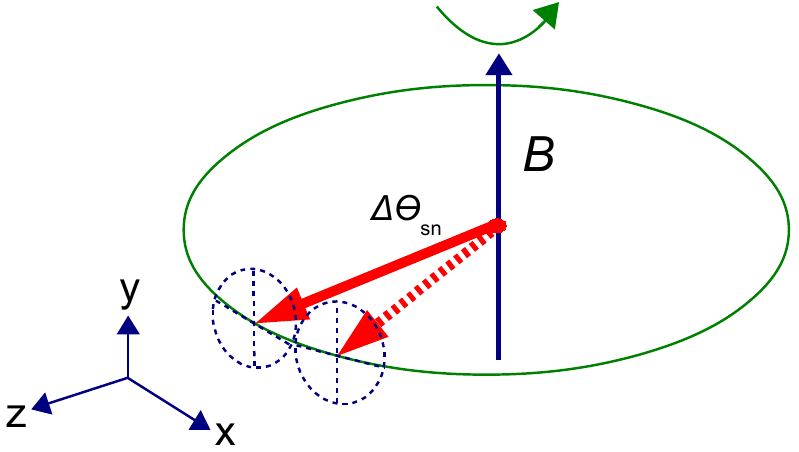} 

\hskip 2.5cm (a)
\vskip0.5cm

\includegraphics[width=8.6cm]{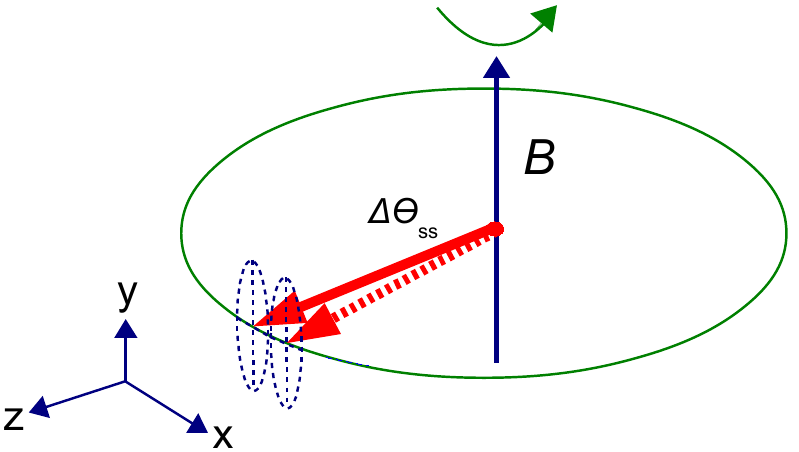}

\hskip 2.5cm(b)
\vskip0.5cm
\caption{
(a) Magnetometry with an ensemble of spins, all pointing into the $z$ direction. (solid arrow) The large collective spin precesses around the magnetic field pointing into the $y$ direction. (dashed arrow) After a precession of an angle $\Delta \theta_{\rm sn},$ the uncertainty ellipse of the spin is not overlapping with the uncertainty ellipse of the spin at the starting position. Hence, $\Delta \theta_{\rm sn}$ is close to the uncertainty of the phase estimation, which coincides with the shot-noise limit. (b) Magnetometry with an ensemble of spins, all pointing to the $z$ direction and spin squeezed along the $x$ direction. The uncertainty of the phase estimation is close to the angle  $\Delta \theta_{\rm ss},$ which is smaller than $\Delta \theta_{\rm sn}$ due to the spin squeezing.   
}
\label{fig:magneto}
\end{center}
\end{figure*}

Let us start with a basic scheme for magnetometry using an almost completely polarized state. 
The total spin of the ensemble, originally pointing into the $z$-direction, is rotated by a magnetic field pointing to the $y$-direction, 
as can be seen in \FIG{fig:magneto}(a). Hence, the unitary giving
the dynamics of the system is
 \be \label{eq:Uy}
 U_{\theta}=e^{-iJ_y\theta}.
 \ee
The stronger the field, the faster the rotation. The rotation angle can be obtained by measuring 
$\exs{J_x},$ a spin component perpendicular to the initial spin. The estimation of rotation angle
in such an experiment is a particular case of Ramsey-interferometry (see, e.g., \REFS{0953-4075-45-10-103001,2014arXiv1405.7703D})  
and has been realised for magnetometry in cold atoms \cite{PhysRevLett.104.133601}.
This idea has been used in experiments with cold gases to get even a spatial information on the magnetic field and its gradient \cite{wildermuth2006sensing,PhysRevLett.98.200801,behbood2013real,koschorreck2011high,2014arXiv1405.6022M}.

So far, it looks as if the mean spin behaves like a clock arm and its position tells us the value of the magnetic field. 
At this point one has to remember that we have an ensemble of particles governed by quantum mechanics, and the uncertainty of the spin component perpendicular to the mean spin is not zero. For the completely polarised state, the squared uncertainty is
\begin{equation}
\va{J_x}=\tfrac{N}{4}. \label{eq:unc}
\end{equation}
Hence, the angle of rotation can be estimated only with a finite precision as can be seen in \FIG{fig:magneto}(a).
Intuitively speaking, we can detect a rotation $\theta$ only if the uncertainty ellipses of the spin in the original position and in the position after the rotation do not overlap with each other too much. 
Based on these ideas and \EQ{eq:unc}, 
with elementary geometric considerations, 
we arrive at
\begin{equation}
(\Delta\theta)^2 \sim \frac{\va{J_x}}{\exs{J_z}^2}=\tfrac{1}{N}. \label{eq:unc2}
\end{equation}
Thus, we obtained the shot-noise scaling \eqref{eq:shotnoise}, even with very simple, qualitative arguments.

A more rigorous argument is based on the formula \eqref{acc}, where we measure the operator 
\be
M=J_x.
\ee 
The expectation value and the variance of this operator, as a function of $\theta,$ are
\bea
\exs{M}(\theta)&=&\exs{J_z}\sin(\theta)+\exs{J_x}\cos(\theta),\nonumber\\
\va{M}(\theta)&=&\va{J_x}\cos^2(\theta)+
\va{J_z}\sin^2(\theta)\nonumber\\
&+&\left(\tfrac{1}{2}\exs{J_xJ_z+J_zJ_x}-\exs{J_x}\exs{J_z}\right)\sin(2\theta).\label{eq:M_fullypolarized}
\eea 
Hence, using \exs{J_x}=0, we obtain for the precision
\begin{equation}
\va{\theta}\vert_{\theta=0} = \frac{\va{J_x}}{\exs{J_z}^2}, \label{eq:accuracy}
\end{equation}
which equals $\frac{1}{N},$ demonstrating a shot-noise scaling
 for the totally polarised states.

We can see that $\va{\theta}$ could be smaller 
if we decrease $\va{J_x}$ \cite{0295-5075-42-5-481}.
 A comparison between \FIG{fig:magneto}(a) and (b)  
 also demonstrates the fact that a smaller $\va{J_x}$ leads to 
a higher precision. 
The variances of the angular momentum
components are bounded by the Heisenberg uncertainty relation \cite{PhysRevA.47.5138}
\begin{equation}
\va{J_x}\va{J_y}\ge \tfrac{1}{4}\vert\exs{J_z}\vert^2.\label{eqKU}
\end{equation}
Thus, the price of decreasing $\va{J_x}$ is increasing $\va{J_y}.$

Let us now characterise even quantitatively the properties of the state that can reach an
improved metrological precision. For fully polarised states, \EQ{eqKU}
is saturated such that 
\be
\va{J_x}=\va{J_y}=\tfrac{1}{2}\vert\exs{J_z}\vert.
\ee
Due to decreasing $\va{J_x}$, our state fulfils
\begin{equation}
\va{J_x} < \tfrac{1}{2}\vert\exs{J_z}\vert,\label{eqKU2}
\end{equation}
where $z$ is the direction of the mean spin, and the bound in \EQ{eqKU2} 
is the square root of the bound in \EQ{eqKU}.
Such states are called {\it spin squeezed states}  \cite{PhysRevA.47.5138,PhysRevA.50.67,PhysRevLett.86.4431,2011PhR...509...89M}.
In practice this means that the
mean angular momentum of the state is large, and in a direction orthogonal to the mean spin
the uncertainty of the angular momentum is small. An alternative and slightly different
definition of spin squeezing considers the usefulness of spin squeezed states
for reducing spectroscopic noise in a setup different from the one discussed in this section 
\cite{PhysRevA.50.67}.
Spin squeezing has been realised in many experiments with cold atomic ensembles.
In some systems the particles do not interact with each other, and light is used for spin squeezing
\cite{0953-4075-45-10-103001,PhysRevLett.104.133601,PhysRevLett.101.073601,PhysRevLett.83.1319,julsgaard2001experimental,RevModPhys.82.1041},
while in Bose-Einstein condensates the spin squeezing can be generated using the inter-particle interaction
\cite{esteve2008squeezing,riedel2010atom,PhysRevLett.111.143001,2014arXiv1405.6022M}.

Next, we can ask, what the best possible
phase estimation precision is for the metrological
task considered in this section. For that, 
we have to use the following inequality based on
general principles of angular momentum theory
\begin{equation}
\label{Jxyz2}
\ex{J_x^2+J_y^2+J_z^2}\le \tfrac{N(N+2)}{4}.
\end{equation}
Note that 
equation \eqref{Jxyz2} is saturated only by symmetric multiqubit states.
Together with the identity connecting the second moments, variances and expectation values
\begin{equation}
\label{varlength}
\va{J_l}+\ex{J_l}^2=\ex{J_l^2},
\end{equation}
\EQLL{Jxyz2} leads to a bound on the uncertainty in the squeezed orthogonal direction
\begin{equation}
\label{eq:Jxy}
\va{J_y}
\le \tfrac{N(N+2)}{4}-\ex{J_z}^2.
\end{equation}
Introducing the maximal spin length 
\be 
J_{\max}=\tfrac{N}{2},
\ee  
we arrive at the inequality
\begin{equation}\label{eq:bnd1}
\va{J_y}
\le \tfrac{N}{2}+\tfrac{N^2}{4}\left(1-\tfrac{\exs{J_z}^2}{J_{\max}^2}\right).
\end{equation}
This leads to a simple bound on the precision 
\begin{equation}\label{eq:bnd2}
(\Delta \theta)^{-2} = \frac{\exs{J_z}^2}{\va{J_x}} \le 4\va{J_y}
\le 2N+N^2\left(1-\tfrac{\exs{J_z}^2}{J_{\max}^2}\right),
\end{equation}
which indicates that the precision is limited for almost completely polarised spin squeezed states with $\exs{J_z}\approx J_{\max}.$
Here, the equality in \EQ{eq:bnd2} is based on \EQ{eq:accuracy}, while the first inequality is due
to \EQ{eqKU}, and the second one comes from \EQ{eq:bnd1}.
The bound in  \EQ{eq:bnd2}  is not optimal, as for the fully polarised state we would expect $(\Delta \theta)^{-2} =N,$  while
 \EQ{eq:bnd2}  allows for a higher precision for $\exs{J_z}= J_{\max}.$ 
 
\begin{figure}
\begin{center}
\includegraphics[width=7.6cm]{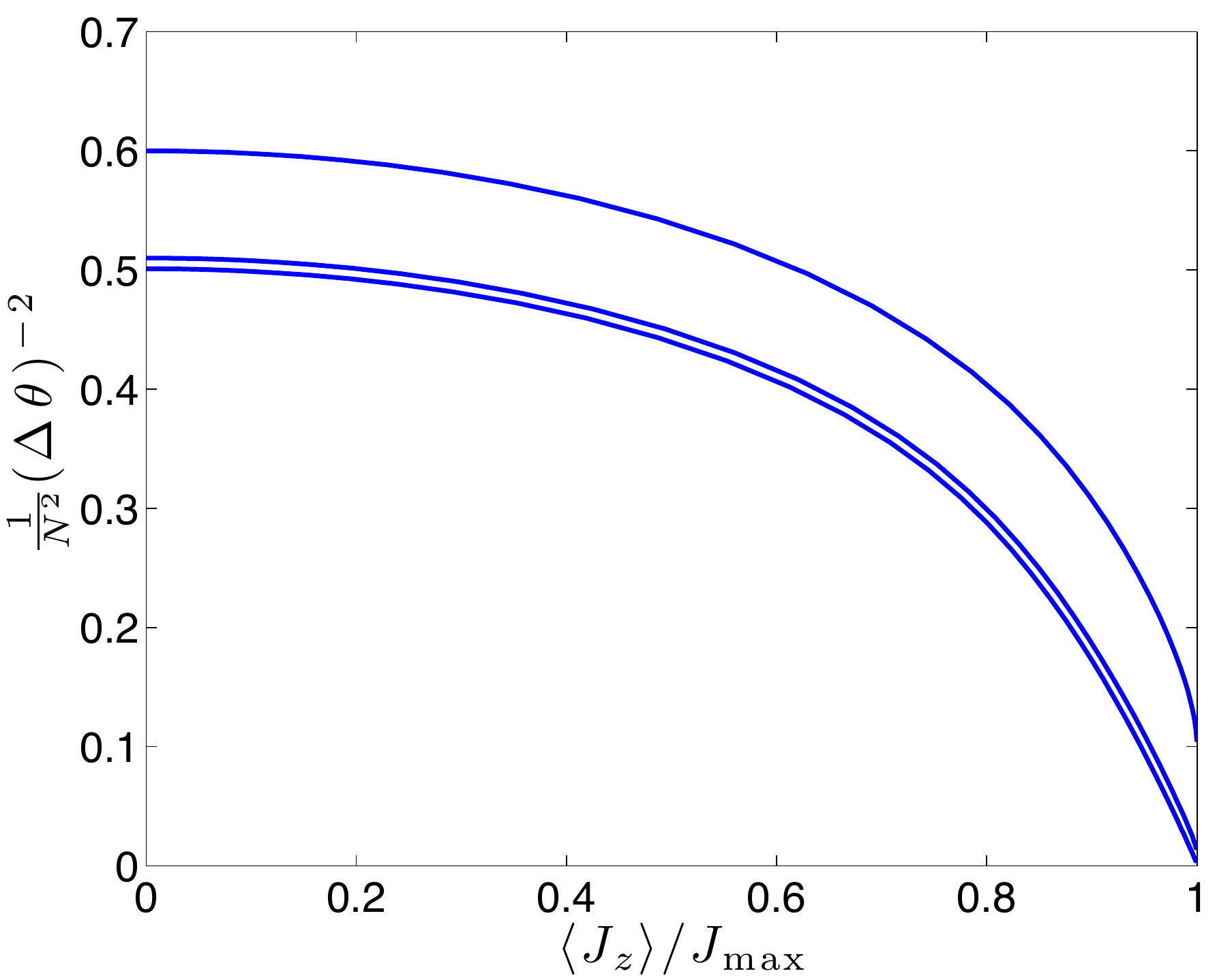} 
\caption{
The best precision achievable  \eqref{eq:accuracy}  divided by $N^2$ as the function of the total spin, from top to bottom, for $N=10,100,1000$ particles. The curves converge to the same curve for large $N,$ which demonstrates an $N^2$ scaling for the precision. Note that the maximum appears in the limit in which the spin length is zero.}
\label{fig:maxsqueezing}
\end{center}
\end{figure}

 It is possible to obtain the best achievable precision numerically for our case, 
 when $\exs{J_x}$ is measured for a state that is almost completely polarised in the $z$-direction
 in a field pointing into the $y$-directon.
For even $N,$ states giving the smallest $\va{J_x}$  for a given $\exs{J_z}$ can be obtained as a ground state of the Hamiltonian \cite{PhysRevLett.86.4431}
\begin{equation}\label{HL}
H(\Lambda)=J_x^2-\Lambda J_z,
\end{equation}
where $\Lambda\ge 0$ plays the role of a Lagrange multiplier.
This also means that the ground states of $H(\Lambda)$ give the best $(\Delta \theta)^{-2}$ for a given $\exs{J_z},$
when collective operators are measured for estimating $\theta.$
Since the ground state of \EQ{HL} is symmetric, it is possible to make the calculations in the symmetric subspace
and hence model large systems.
We plotted the precision $(\Delta \theta)^{-2}$ as a function of the polarisation $\exs{J_z}$ for different values of $N$  in \FIG{fig:maxsqueezing},
which demonstrates that $(\Delta \theta)^{-2}$ scales as $N^2.$
Hence,  for the precision of phase-estimation the Heisenberg scaling \eqref{eq:Heisenberg} can be reached.

Paradoxically the maximum is reached in the limit of zero mean spin.
An added noise can radically change this situation. If the mean spin is small, 
and its direction is the information that we use for metrology, then a very small added noise
can change the direction of the spin, making the metrology for this case
impractical. Thus, if we consider local noise acting on each particle independently, 
then the maximum $(\Delta \theta)^{-2}$ will be reached
at a finite spin length.

\subsection{Metrology with a GHZ state}
\label{sec:Metrology with a GHZ state}

Next, we will show another example where the Heisenberg scaling
for the precision of phase estimation can be reached.
The scheme is based on a GHZ state defined as 
\begin{equation}\label{eq:GHZ}
\ket{{\rm GHZ}_N}=\tfrac{1}{\sqrt{2}}(\ket{0}^{\otimes N}+\ket{1}^{\otimes N}),
\end{equation}
where we follow the usual convention defining the $\ket{0}$ and $\ket{1}$ states with the eigenstates of $j_z$ as 
$\ket{0}\equiv\ket{+\tfrac{1}{2}}_z$ and $\ket{1}\equiv\ket{-\tfrac{1}{2}}_z.$ 
Such states have been created in photonic systems \cite{ThreeQubitGHZ1,ThreeQubitGHZ,FourPhotonGHZ,lu2007experimental,gao2010experimental}  and in cold trapped ions \cite{leibfried2004toward,sackett2000experimental,PhysRevLett.106.130506}.
Let us consider the dynamics given by 
\be U_{\theta}=e^{-iJ_z\theta}.
\ee
Under such dynamics, the GHZ state evolves as 
\begin{equation}
\ket{{\rm GHZ}_N}(t)=\tfrac{1}{\sqrt{2}}(\ket{0}^{\otimes N}+e^{-iN\theta}\ket{1}^{\otimes N}),
\end{equation}
hence the difference of the phases of the two terms scales as $\sim N.$
Let us consider measuring the operator
\begin{equation}\label{eq:corrop}
M=\sigma_x^{\otimes N},
\end{equation}
which is essentially the parity in the $x$-basis.
Note that this operator needs an individual access to the particles.
For the dynamics of the expectation value and the variance we obtain
\begin{equation}\label{eq:MMM}
\exs{M}=\cos(N \theta),\;\;\;\;\;\;\;\;\va{M}=\sin^2(N \theta).
\end{equation}
Hence, based on \EQ{acc}, for small $\theta$ the precision is
\begin{equation}
(\Delta\theta)^2\vert_{\theta=0}=\tfrac{1}{N^2},
\end{equation}
which means that we reached the Heisenberg scaling \eqref{eq:Heisenberg}.
 In \REF{leibfried2004toward}, the scheme described above has been realised experimentally with three ions and a precision above the shot-noise limit has been achieved.

Note, however, that the GHZ state is very sensitive to noise.
Even if a single particle is lost, it becomes a separable state.
Thus, it is a very important question, how well such a state 
can be created, and how noise is influencing
the scaling of the precision with the particle number.
This question will be discussed in \SEC{Sec:Metrology with noise}.
Concerning spin squeezed states and GHZ states, it has been observed that
under local noise, such as dephasing and particle loss, for large particle numbers, the GHZ state becomes
useless while the spin squeezed states, discussed in the previous section, are optimal \cite{1367-2630-15-7-073043}. 

A related metrological scheme for two-mode systems is based on a Mach-Zender interferometer \cite{PhysRevD.23.1693,PhysRevD.30.2548,PhysRevLett.69.3598,PhysRevLett.71.1355,PhysRevLett.75.2944,PhysRevLett.110.163604,PhysRevA.88.042316},
using as inputs NOON states defined as \cite{2008ConPh..49..125D}
\begin{equation}\label{eq:NOON}
\ket{{\rm NOON}}=\tfrac{1}{\sqrt{2}}(\ket{N,0}+\ket{0,N}).
\end{equation}
Here, the state $\ket{n_1,n_2}$ describes a system with $n_1$  particles in the first bosonic mode and $n_2$  particles in the second bosonic mode. 
For example, the two modes can be two optical modes, or, two spatial modes in a double-well potential.
Thus, similarly to GHZ states, the state is a superposition of two states: all particles in the first state and
all particles in the second state. However, in this scheme 
we do not have a local access to the particles.
Hence, we cannot easily measure the operator \eqref{eq:corrop}, 
which is a multi-particle correlation operator, and  instead
the following operator has to be measured
\begin{equation}
M=\ket{N,0}\bra{0,N}+\ket{0,N}\bra{N,0}.
\end{equation}
The basic idea of the $N$-fold gain in precision is similar to the idea used for the method based on the GHZ state.
The expectation value of $M$ and the variance of $M$ as a function of $\theta$ is the same 
as before, given in \EQ{eq:MMM}.
With that, the Heisenberg scaling can be reached. 
Metrological experiments with NOON states have been carried out in optical systems that surpassed the shot-noise limit
 \cite{mitchell2004super,PhysRevLett.94.090502,nagata2007beating,PhysRevLett.98.223601}.

\subsection{Metrology with a symmetric Dicke state}
\label{Sec:Metrology with a symmetric Dicke state}

As a third example, we will consider metrology with $N$-qubit symmetric Dicke states
\begin{equation}\label{eq:DickeNm}
\ket{D_N^{(m)}}=\binom{N}{m}^{-\frac{1}{2}}\sum_k \mathcal{P}_k (\ket{1}^{\otimes m}\otimes\ket{0}^{\otimes (N-m)}),
\end{equation}
where the summation is over all the different permutations of $m$ $1$'s and $(N-m)$ $0$'s.
One of such states is the $W$-state for which $m=1,$ which has been prepared with photons and ions \cite{PhysRevLett.92.077901,haffner2005scalable}. 

\begin{figure}
\begin{center}
\includegraphics[width=8.6cm]{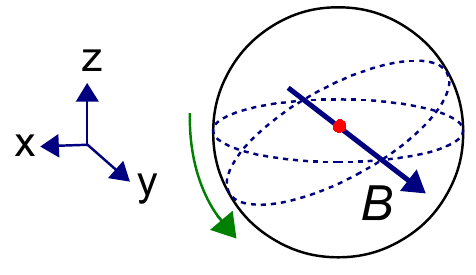}
\caption{
Magnetometry with Dicke states of the form \eqref{eq:Dicke}. The uncertainty ellipse of the state is rotated around the magnetic field pointing to the $x$-direction. The rotation angle can be estimated by measuring the uncertainty of the $z$-component of the collective spin. Note that, unlike in the case of a fully polarised state, after a rotation of an angle $\pi$ we obtain the original Dicke state. }
\label{fig:dicke}
\end{center}
\end{figure}

From the point of view of metrology, we are interested mostly in the symmetric Dicke state for even $N$ and $m=\frac{N}{2}.$ 
This state is known to be highly entangled \cite{Toth:07}.
In the following, we will omit the superscript giving the number of  $1$'s and use the notation
\begin{equation}\label{eq:Dicke}
\ket{D_N}\equiv\ket{D_N^{(\frac{N}{2})}}.
\end{equation}
Symmetric Dicke states of the type \eqref{eq:Dicke} have been created in photonic systems \cite{PhysRevLett.98.063604,PhysRevLett.103.020504,PhysRevLett.103.020503,PhysRevLett.107.080504,PhysRevLett.109.173604}
and in cold gases 
\cite{lucke2011twin,hamley2012spin,PhysRevLett.112.155304}.
In \REFS{lucke2011twin,PhysRevLett.107.080504}, their metrological properties have also been verified. 

The state \eqref{eq:Dicke} has $\exs{J_l}=0$ for all $l=x,y,z.$ 
For the second moments we obtain
\begin{equation}\label{eq:DickePar}
\exs{J_x ^2}=\exs{J_y ^2}=\tfrac{N(N+2)}{8},\;\;\;\;\;\;\;\exs{J_z ^2}=0.
\end{equation}
It can be seen that  $\exs{J_z ^2}$ is minimal,  $\exs{J_x ^2}$ and  $\exs{J_y ^2}$
are close to the largest possible value, $\frac{N^2}{4}.$

The state has a rotational symmetry around the $z$ axis.
Thus,  the state is not changed by dynamics of the type
$\exp(-iJ_z\theta).$ 
Based on these considerations, we will use dynamics of the type \eqref{eq:Uy}.
Moreover, since the total spin length is zero, a rotation around any axis remains
undetected if we measure only the expectation values of the collective angular momentum components.
Hence, our setup will measure the expectation value of
\begin{equation}
M=J_z^2.
\end{equation}
Note that this is also a collective measurement. In practice, to measure $\exs{M},$ we have to measure $J_z$ many times and 
compute the average of the squared values.

For the dynamics of the expectation value we obtain
\begin{equation} \label{Mdicke}
\exs{M}=\tfrac{N(N+2)}{8}\sin^2( \theta)\equiv\tfrac{N(N+2)}{8}\frac{1-\cos(2\theta)}{2}.
\end{equation}
The expectation value $\exs{J_z^2}$  starts from zero, and
 oscillates with a frequency twice as large as the frequency of the oscillation was for
the analogous case for $\exs{J_x}$ in \EQ{eq:M_fullypolarized}.
This is due to fact that  after a rotation with an angle $\pi$ we obtain again the original Dicke state.
Following the calculations given in \REF{lucke2011twin}, we arrive at
\begin{equation}
(\Delta\theta)^{2}\vert_{\theta=0}=\tfrac{2}{N(N+2)},
\end{equation}
which again means that we reached the Heisenberg scaling \eqref{eq:Heisenberg}.
The quantum dynamics of the Dicke state used for metrology is depicted in \FIG{fig:dicke}.

\subsection{Singlet states}
\label{sec:Singlet states}

\begin{figure}
\centerline{ \epsfxsize14cm \epsffile{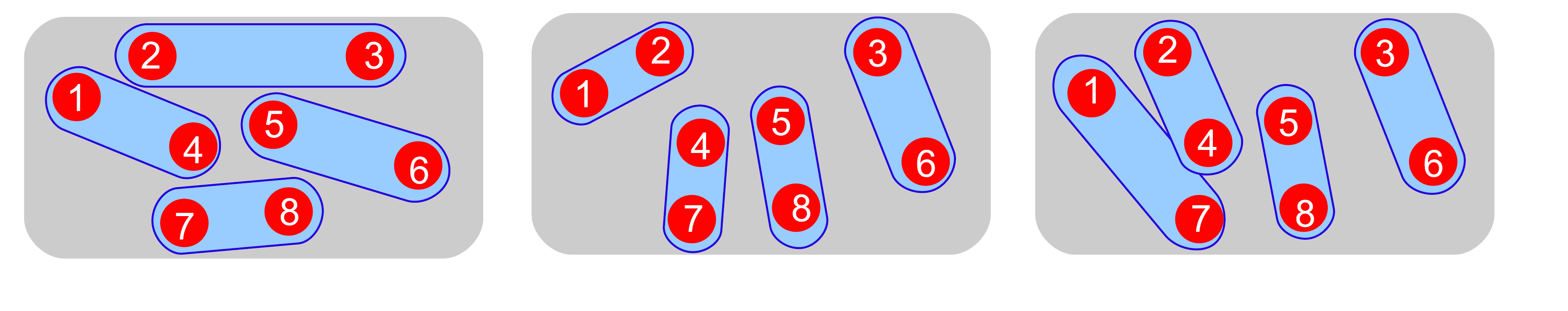}}
\caption{
For spin-$\frac{1}{2}$ particles, the permutationally invariant singlet is an equal mixture of all possible arrangements of two-particle singlets.
Three of such arrangements are shown for eight particles. Note that the eight atoms are arranged in the same way on the figures,
only the pairings are different.
 }
 \label{fig:singlet}
\end{figure}

Finally, we show another example for states that can be used for metrology
in large particle ensembles. Pure singlet states are simultaneous eigenstates of $J_l$ for $l=x,y,z$ with an eiganvalue zero, that is,
\begin{equation}\label{eq:singletdef}
J_l\ket{\Psi_{\rm s}}=0.
\end{equation}
Mixed singlet states are mixtures of pure singlet states, and hence $\ex{J_{\vec n}^m}=0$ for any direction $\vec n $ and any power $m.$
Such states can be created in cold atomic ensembles by squeezing the uncertainties of all the three collective spin components \cite{1367-2630-12-5-053007,2013PhRvL.111j3601B,2014arXiv1403.1964B}. Since in large systems practically all initial states and all the possible dynamics are permutationally invariant,
they are expected to be also permutationally invariant. For spin-$\frac{1}{2}$ particles, there is a unique
 permutationally invariant singlet state \cite{PhysRevA.88.013626}
\begin{equation}
\label{eq:singlet}
\varrho_{\rm s} =\frac{1}{N!}\sum_k \Pi_k ( \ket{\singlet} \bra{\singlet} 
\otimes \cdots \otimes \ket{\singlet} \bra{\singlet}) \Pi_k^\dagger,
\end{equation}
where the summation is over all permutation operators $\Pi_k$ and 
\begin{equation}\label{singlet}
\ket{\singlet}=\frac{1}{\sqrt{2}}\left(\ket{01}-\ket{10}\right).
\end{equation}
A realisation of the singlet state \eqref{eq:singlet} with an ensemble of particles is shown in \FIG{fig:singlet}.

A singlet state is invariant under $\exp(-iJ_{\vec n}\theta)$ for any $\vec n.$ 
Thus, it is completely insensitive to rotations
around any axis. How can it be useful for magnetometry?
Let us now assume that we would like to analyse a magnetic field pointing in the $y$-direction using
spins placed in an equidistant chain.  While the singlet \eqref{eq:singlet} is insensitive to the 
homogenous component of the magnetic field, 
it is very sensitive to the dynamics 
\begin{equation}
e^{-i\sum_n n j_y^{(n)}\theta_G},
\end{equation}
where $\theta_G$ is proportional to the field gradient.
This makes the state useful for differential magnetometry, 
since singlets are insensitive to external homogeneous magnetic fields, while sensitive to the 
gradient of the magnetic field \cite{PhysRevA.88.013626}.
Similar ideas work even if the atoms are in a 
cloud rather than in a chain.
The quantity to measure in order to estimate $\theta_G$ is again 
$\exs{J_z^2}$ as was the case in \SEC{Sec:Metrology with a symmetric Dicke state}.
This idea is also interesting even for a bipartite singlet of two large spins \cite{PhysRevLett.105.013603}.

\section{Spin squeezing and entanglement}
\label{sec:Spin squeezing and entanglement}

As we have seen in \SEC{Sec:Spin squeezed states}, spin squeezed states have been more useful for metrology
than fully polarised product states. Moreover, states very different from product states, such as GHZ states
and Dicke sates could reach the Heisenberg limit in parameter estimation. Thus, 
large quantum correlations, or entanglement, can help in metrological tasks.
In this section, we will discuss some relations between entanglement and spin squeezing,
showing why entanglement is necessary to surpass the shot-noise limit.
We also discuss that not only entanglement, but true multipartite entanglement is needed
to reach the maximal precision in the metrology with spin squeezed states.

\subsection{Entanglement and multi-particle entanglement}
\label{sec:Entanglement and multi-particle entanglement}

Next, we need the following definition. A quantum state is (fully) separable if it can be written as \cite{PhysRevA.40.4277}
\begin{equation}\label{eq:sep}
\varrho_{{\rm sep}}=\sum_m p_m \rho_m^{(1)} \otimes \rho_m^{(2)} \otimes...\otimes\rho_m^{(N)},
\end{equation}
where $\rho_m^{(n)}$ are single-particle pure states.
Separable states are essentially states that can be created without an inter-particle
interaction, just by mixing product states.
States that are not separable are called entangled.
Entangled states are more useful than separable ones
for several quantum information processing tasks, such as quantum teleportation,
quantum cryptography, and, as we will show later, for quantum metrology \cite{RevModPhys.81.865,guhne2009entanglement}. 

In the many-particle case, it is not sufficient to distinguish only two qualitatively
different cases of separable and entangled states.
For example, an $N$-particle state is entangled,
even if only  two of the particles are entangled with each other, 
while the rest of the particles are, say, in the  state $\ket{0}.$ Usually,
such a state we would not call multipartite entangled.
This type of entanglement is very different from the entanglement of
a GHZ state \eqref{eq:GHZ}.

Hence, the notion of genuine multipartite entanglement \cite{sackett2000experimental,PhysRevLett.87.040401} has been
introduced to distinguish partial entanglement from the case when
all the particles are entangled with each other. It is defined as follows.
A pure state is biseparable, if it can be written
as a tensor product of two multi-partite states
\begin{equation}
\ket{\Psi}=\ket{\Psi_1}\otimes \ket{\Psi_2}.
\end{equation}
A mixed state is biseparable if it can be written as a mixture of biseparable
pure states.  A state that is not biseparable, is genuine multipartite entangled.
In many quantum physics experiments the goal was to create 
genuine multipartite entanglement, as this could be used to demonstrate
that something qualitatively new has been created compared to
experiments with fewer particles \cite{ThreeQubitGHZ1,ThreeQubitGHZ,FourPhotonGHZ,lu2007experimental,gao2010experimental,sackett2000experimental,PhysRevLett.98.063604,PhysRevLett.103.020504,PhysRevLett.103.020503,PhysRevLett.109.173604,PhysRevLett.106.130506}.

In the many-particle scenario, further levels of multi-partite
entanglement must be introduced as verifying full $N$-particle entanglement
for $N=1000$ or $10^6$ particles is not realistic.
In order to characterise the different levels of multipartite entanglement,
we start first with pure states. We call a state $k$-producible, if it can be
written as a tensor product of the form
\begin{equation}
\ket{\Psi}=\otimes_m \ket{\psi_m},
\end{equation}
where $\ket{\psi_m}$ are multiparticle states with at most $k$ particles.
A $k$-producible state can be created in such a way
 that only particles within groups containing
not more than  $k $ particles were interacting with each other.
This notion can be extended to mixed states by calling
a mixed state $k$-producible if it can be written as a mixture of
pure $k$-producible states. A state that is not $k$-producible contains at least 
$(k+1)$-particle entanglement \cite{1367-2630-7-1-229,PhysRevA.73.052319}. 
Using another terminology, we can also say that the entanglement depth
of the quantum state is larger than $k$ \cite{PhysRevLett.86.4431}.

\begin{figure}
\begin{center}
\includegraphics[width=11cm]{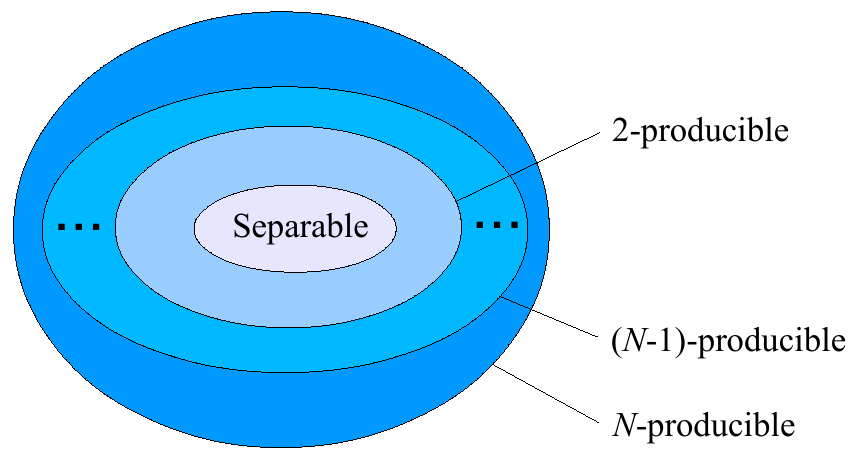}
\caption{
Sets of states with various forms of multipartite entanglement. 
$k$-producible states form larger and larger convex sets, $1$-producible states being equal
to the set of separable states, while the set of physical quantum states is equal to the set of $N$-producible states.
\label{fig:set}}
\end{center}
\end{figure}

It is instructive to depict states with various forms of multipartite entanglement
in set diagrams as shown in \FIG{fig:set}. Separable states are a convex set since
if we mix two separable states, we can obtain only a separable state.
Similarly, $k$-producible states also form a convex set.
In general, the set of $k$-producible states contain the set of $l$-producible states
if $k>l.$ 

\subsection{The original spin squeezing criterion}
\label{sec:The original spin squeezing criterion}

Let us see now how entanglement and multiparticle entanglement is related to spin squeezing.
It turns out that spin squeezing, discussed in \SEC{Sec:Spin squeezed states}, is strongly related to
entanglement.
A ubiquitous entanglement criterion in this context is
the spin squeezing inequality
 \cite{sorensen2001many}
\begin{eqnarray}
\xi_{\rm s}^2:=N\frac{\va{J_x}}{\exs{J_y}^2+\exs{J_z}^2}\ge 1.
\label{motherofallspinsqueezinginequalities}
\end{eqnarray}
If a state violates
\EQ{motherofallspinsqueezinginequalities},
then it is entangled (i.e., not fully separable).
In order to violate \EQ{motherofallspinsqueezinginequalities},
its denominator must be large while its numerator must be small,
hence, it detects states that have 
a large spin in some direction, while a small variance of a spin component 
in an orthogonal direction.
That is, \EQ{motherofallspinsqueezinginequalities} detects the entanglement of spin
squeezed states depicted in \FIG{fig:magneto}(b).

For spin  squeezed states, it has also been noted that
multipartite entanglement, not only simple nonseparability is needed
for large spin squeezing \cite{PhysRevLett.86.4431}. 
To be more specific, for a given mean spin length, larger and larger spin squeezing
is possible only if the state has higher and higher levels of multipartite entanglement.
Moreover, larger and larger spin squeezing leads to larger and larger measurement
precision.
Such strongly squeezed states have been created experimentally in cold gases
and a 170-particle entanglement has been detected \cite{gross2010nonlinear}.

At this point note that only the first and second moments of the collective quantities are needed to evaluate 
the spin squeezing condition \eqref{motherofallspinsqueezinginequalities}. It is easy to 
show that all these can be obtained from the average two-particle density matrix of the quantum state defined as
\cite{PhysRevA.79.042334}
\be
\varrho_{{\rm av}2}=\frac{1}{N(N-1)}\sum_{m\ne n }\varrho_{mn},
\ee
where $\varrho_{mn}$ is the reduced two-particle state of particles $m$ and $n.$

In summary, entangled states seem to be more useful than separable ones
for magnetometry with spin squeezed states discussed in \SEC{Sec:Spin squeezed states}. 
Moreover, states with $k$-particle entanglement
can be more useful than states with $(k-1)$-particle entanglement for the same 
metrological task. This finding will be extended to general metrological tasks in \SEC{sec:Quantum Fisher information and entanglement}.

\subsection{Generalised spin squeezing criteria}
\label{sec:Generalised spin squeezing criteria}

The original spin squeezing entanglement criterion \eqref{motherofallspinsqueezinginequalities} 
can be used to
detect the entanglement of almost completely polarised spin squeezed states.
However, there are other highly entangled states, such as Dicke states \eqref{eq:Dicke} and singlet states defined in \eqref{eq:singletdef}.
For these states, the denominator of the fraction in  \EQ{motherofallspinsqueezinginequalities} is zero,
thus they are not detected by the original squeezing entanglement criterion.

A complete set of entanglement conditions similar to the condition \eqref{motherofallspinsqueezinginequalities} has been determined, called the optimal spin squeezing inequalities. They are called optimal since, in the large particle number limit, 
they detect all entangled states that can be detected based on
the first and second moments of collective angular momentum components.
For separable states of the form \eqref{eq:sep}, the following inequalities are satisfied 
\cite{PhysRevA.79.042334}
\begin{subequations}\label{Jxyzineq}
\begin{eqnarray}
\exs{J_x^2}+\exs{J_y^2}+\exs{J_z^2} &\le& \tfrac{N(N+2)}{4},
\label{theorem1a}
\\
\va{J_x}+\va{J_y}+\va{J_z} &\ge& \tfrac{N}{2},
\label{Jxyzineq_singlet}
\\
\exs{J_k^2}+\exs{J_l^2}-\tfrac{N}{2} &\le&  (N-1)\va{J_m},
\label{Jxyzineq_spsq2}\\
(N-1)\left[\va{J_k}+\va{J_l}\right] &\ge&
\exs{J_m^2}+\tfrac{N(N-2)}{4}, \label{Jxyzineq_spsq3} \;\;\;\;\;\;
\end{eqnarray} 
\end{subequations}
where $k,l,m$ take all the possible permutations of $x,y,z.$
The inequality \EQ{theorem1a}, identical to \EQ{Jxyz2}, is valid for all quantum states.
On the other hand, violation of any of the inequalities
(\ref{Jxyzineq}b-d)  implies
entanglement.

Based on the entanglement conditions \eqref{Jxyzineq}, new spin squeezing parameters
have been defined. 
For example, \EQ{Jxyzineq_spsq2} is equivalent to \cite{PhysRevA.89.032307}
\begin{eqnarray}
\xi_{\rm os}^2:=(N-1)\frac{\va{J_x}}{\exs{J_y^2}+\exs{J_z^2}-\frac{N}{2}}\ge 1,
\label{xi_os}
\end{eqnarray}
provided that the denominator of \eqref{xi_os} is positive. The criterion \eqref{xi_os} 
can be used to detect entanglement close to Dicke states, discussed in \SEC{Sec:Metrology with a symmetric Dicke state}. One can see that for the Dicke state \eqref{eq:Dicke}, 
the numerator of the fraction in \EQ{xi_os} is zero,
while the denominator is maximal [see \EQ{eq:DickePar}]. 
Apart from entanglement, it is also possible to detect multiparticle entanglement close to Dicke states.
A condition linear in expectation values and variances of collective observables 
has been presented in \REF{PhysRevLett.107.180502} for detecting multipartite entanglement.
 A nonlinear criterion is given in 
\REF{PhysRevLett.112.155304}, which detects all states as multipartite entangled that can be 
detected based on the measured quantities.
The criterion has  been used even experimentally \cite{PhysRevLett.112.155304}.
An entanglement depth of $28$ particles has been detected in an ensemble of around $8000$ cold atoms.

The inequality \EQ{Jxyzineq_singlet} is equivalent to \cite{1367-2630-12-5-053007,2014arXiv1403.1964B}
\begin{eqnarray}
\xi_{\rm singlet}^2:=\frac{\va{J_x}+\va{J_y}+\va{J_z}}{\frac{N}{2}}\ge 1.
\label{xi_singlet}
\end{eqnarray}
 The parameter $\xi_{\rm singlet}^2$ can be used to detect entanglement close to singlet states discussed in \SEC{sec:Singlet states}. It can be shown that the number of non-entangled spins in the ensemble
 is bounded from above by $N\xi_{\rm singlet}^2.$

Finally, it is interesting to ask, what the relation of the new spin squeezing parameters is to the original one. It can be proved that  the parameters $\xi_{\rm singlet}^2$ and $\xi_{\rm os}^2$ detect all entangled states that are detected by $\xi_{\rm s}^2.$ 
They detect even states not detected by $\xi_{\rm s}^2,$ such as entangled states with a zero mean spin, like Dicke states and singlet states. Moreover, it can be shown that for large particle numbers, $\xi_{\rm os}^2$ in itself is also strictly stronger than $\xi_{\rm s}^2$ \cite{PhysRevA.89.032307}. 

\section{Quantum Fisher information}
\label{sec:The quantum Fisher information}

In this section, we review the theoretical background of quantum metrology, such as the
Fisher information, the Cram\'er-Rao bound and the quantum Fisher information.

\subsection{Classical Fisher information}

Let us consider the problem of estimating a parameter $\theta$ based on measuring a quantity $M.$ Let us assume that the relationship between the two is given by a probability density function $f(x;\theta).$ This function, for every value of the parameter $\theta,$  gives a probability distribution for the $x$ values of $M.$

Let us now construct an estimator $\hat{\theta}(x),$ which would give for every  value $x$ of $M$ an estimate for $\theta.$ In general it is 
not possible to obtain the correct value for $\theta$ exactly. We can still require that the estimator be unbiased, 
that is, the expectation value of  $\hat{\theta}(x)$ should be equal to $\theta.$
This can be expressed as
\be\label{eq:unbiased}
0=\int (\theta- \hat{\theta}(x)) f(x;\theta) dx.
\ee
How well the estimator can estimate $\theta$? The Cram\'er-Rao bound provides a lower bound on the variance of the unbiased estimator as
\be\label{CR}
{\rm var}(\hat{\theta})\ge \frac{1}{F(\theta)},
\ee
where the Fisher information is defined with the probability distribution function $f(x;\theta)$ as 
\be\label{eq:Fisher}
F(\theta)=\int \left(\frac{\partial}{\partial\theta}\log f(x;\theta)\right)^2 f(x;\theta) dx.
\ee
The inequality \eqref{CR} is a fundamental tool in metrology that 
appears very often in physics and engineering, and 
can even be generalised to the case of quantum measurement.
Finally, note that the inequality \eqref{CR} is giving a lower bound
for parameter estimation in the vicinity of a given $\theta.$
This is the {\it local} approach discussed in  
\SEC{sec:Examples}.

\subsection{Quantum Fisher information}

\label{Sec:Quantum Fisher information}

\begin{figure}
\begin{center}
\includegraphics[width=8.6cm]{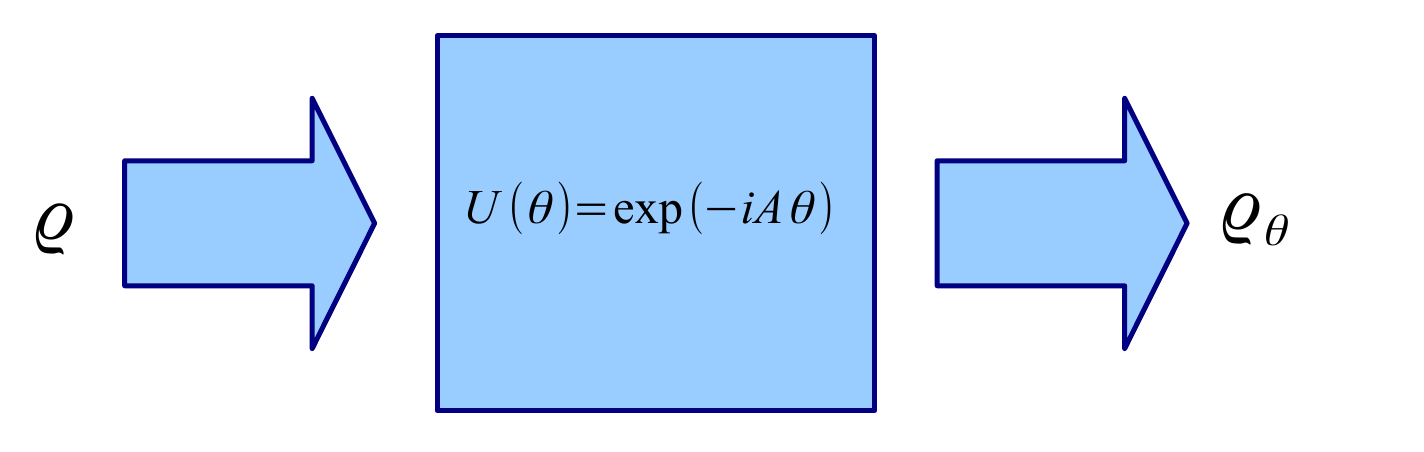}
\caption{
The basic problem of linear interferometry. The parameter
$\theta$ must be estimated by measuring $\varrho_{\theta}.$
\label{fig:interfero}}
\end{center}
\end{figure}

In quantum metrology, as can
be seen in \FIG{fig:interfero}, one of the basic tasks is phase
estimation connected to the unitary dynamics of a linear interferometer
\begin{equation}
\varrho_{\theta}=e^{-i\theta A}\varrho e^{+i\theta A},
\end{equation}
where $\varrho$ is the input state, $\varrho_{\theta}$
is the output state, and $A$ is a Hermitian operator. The operator $A$ can be, for example, 
 a component of the collective
angular momentum $J_l.$ The important question
is, how well we can estimate the small angle $\theta$ by measuring $\varrho_{\theta}.$

Let us use the notion of Fisher information to quantum measurements assuming that 
the estimation of $\theta$ is done based on measuring the operator $M.$ 
Let us denote the projector corresponding to a given measured
value $x$ by $\Pi_x.$ 
Then, we can write
\be\label{eq:f}
f(x;\theta)={\rm Tr}(\varrho_{\theta} \Pi_x),
\ee
which can be used to define an unbiased estimator based on \EQ{eq:unbiased}.
Then, the Fisher information can be obtained using \EQ{eq:Fisher}.
Finally, using  \EQ{eq:f}, the Cram\'er-Rao bound \eqref{CR} gives a lower bound on the precision of the estimation.
A similar formalism works even if the measurements are not projectors,
but in the more general case, positive operator valued measures (POVM).

We could calculate a bound for the precision of the estimation for given
dynamics and a given operator to be measured using this formalism.
However, it might be difficult to find the operator that leads to the best estimation precision
just by trying several operators.
Fortunately, it is possible
to find an upper bound on the precision of the parameter estimation that is valid
for any choice of the operator.
The phase estimation sensitivity, assuming
\emph{any} type of measurement, is limited by the quantum Cram\'er-Rao
bound as \cite{helstrom1976quantum,holevo1982probabilistic}
\begin{equation}\label{eq:CR}
(\Delta\theta)^2\ge\frac{1}{{F_{Q}[\varrho,A]}},
\end{equation}
where $F_{Q}$ is the quantum Fisher information. As a consequence, based on \EQ{acc}, for any operator $M$
we have 
\begin{equation}
\frac{(\Delta M)^2}{\vert \partial_\theta \exs{M}\vert^2} \ge \frac{1}{F_{Q}[\varrho,A]}.
\end{equation}

The quantum Fisher information $F_Q$ can be computed easily with a closed formula.
Let us assume that  a density matrix is given in its eigenbasis as 
\begin{equation}
\varrho=\sum_k \lambda_k \ketbra{ k} .\label{decomp_eig}
\end{equation}
Then, the quantum Fisher information is given as \cite{helstrom1976quantum,holevo1982probabilistic,PhysRevLett.72.3439,braunstein1996generalized}
\begin{equation}
\label{eq:FQ}
F_{Q}[\varrho,A]=2\sum_{k,l}\frac{(\lambda_{k}-\lambda_{l})^{2}}{\lambda_{k}+\lambda_{l}}\vert \langle k \vert A \vert l \rangle \vert^{2}.\label{eq:qF}
\end{equation}

Next, we will review some fundamental properties of the quantum Fisher information which relate it to the variance.

(i) For pure states, from \EQ{eq:FQ} follows
\begin{equation}
\label{eq_ure}
F_{Q}[\varrho,A]=4\va{A}. \label{pure}
\end{equation}

(ii) For all quantum states, it can be proven that
\begin{equation}\label{eq:Fvar}
F_{Q}[\varrho,A]\le4\va{A}.
\end{equation}
This provides an easily computable upper bound on the quantum Fisher information.

For quantum states with $\va{A}=0$ we obtain that $F_{Q}[\varrho,A]=0.$
Such a state does not change under unitary dynamics of the type $e^{-iA\theta}.$
It is instructive to consider the example when $A=J_l$ for $l\in{x,y,z}.$
Then, $\va{J_l}=0$ also implies that the state does not change under the dynamics $e^{-iJ_l\theta},$
as we could see in the case of the singlet states in \SEC{sec:Singlet states}.

(iii) More generally, the quantum Fisher information is convex in the state, that is 
\begin{equation}
F_{Q}[p\varrho_1+(1-p)\varrho_2,A]\le p F_{Q}[\varrho_1,A]+(1-p)F_{Q}[\varrho_2,A].
\end{equation}

(iv) Recently, it has turned out that the quantum Fisher information is the largest convex function that fulfils (i) \cite{PhysRevA.87.032324,yu2013quantum}.
This can be stated in a concise form as follows. Let us consider a very general decomposition of the density matrix
\begin{equation}
\varrho=\sum_k p_k \ketbra{ \Psi_k}, \label{decomp}
\end{equation}
where  $p_k>0$ and $\sum_k p_k=1.$
With that, the quantum Fisher information can be given as the convex roof of the variance,
\begin{equation}
F_{Q}[\varrho,A] =4 \inf_{\{ p_k,\ket{\Psi_k}\}} \sum_k p_k \va {A
 }_{\Psi_k},\label{convroof}
\end{equation}
where the optimisation is over all the possible decompositions \eqref{decomp}.

At this point we have to note that if $\ket{ \Psi_k}$ in the 
decomposition \eqref{decomp} were pairwise orthogonal to each other, then the decomposition \eqref{decomp} 
 would be an eigendecomposition. For density matrices with a non-degenerate spectrum, it would even be unique and easy to obtain by any computer program that diagonalises matrices.
 However, the pure states $\ket{ \Psi_k}$ are {\it not}
required to be pairwise orthogonal, which leads to an infinite number of possible decompositions.
Convex roofs over all such decompositions
appear often in  quantum information science \cite{RevModPhys.81.865,guhne2009entanglement}
in the definitions of  entanglement measures,
for example, the entanglement of formation \cite{PhysRevLett.78.5022,PhysRevLett.80.2245}.
These measures can typically be computed only for small systems. Here, surprisingly, we have the quantum Fisher information  given by a convex roof that can also be obtained as a
 closed formula \eqref{eq:FQ} for any system sizes.

There are generalised quantum Fisher informations different from the original one \eqref{eq:qF}.
 They are convex and have the same value for pure states
as the quantum Fisher information does \cite{PhysRevA.87.032324}. However, they cannot be larger than the quantum Fisher information.
This is counterintuitive: the quantum Fisher information is defined with an infimum, still it is easy to show that
it is the largest, rather than the smallest, among the generalised quantum Fisher informations.
As an example, we mention one of the generalised quantum Fisher informations, defined as four times
the Wigner-Yanase skew information given as \cite{Wigner01061963}
\begin{equation}
\label{eq:WY}
I[\varrho,A]={\rm Tr}(A^2\varrho)-{\rm Tr}(A\varrho^{\frac{1}{2}} A\varrho^{\frac{1}{2}}).
\end{equation}
For pure states, $I[\varrho,A]$ equals the variance and it is convex.
There are even other types of generalised quantum Fisher informations.  
\REFSL{0305-4470-35-4-305,petz2008quantum} introduce an entire family
of generalised quantum Fisher informations, together with a family of 
generalised variances.

Analogously to \EQ{convroof}, it can also be proven that the concave roof of the variance is itself \cite{PhysRevA.87.032324}
\begin{equation}
\va{A}_{\varrho}=\sup_{\{ p_k,\ket{\Psi_k}\}} \sum_k p_k \va {A
 }_{\Psi_k}.\label{convroof2}
\end{equation}
Hence, the main statements can be summarised as follows.
For any decomposition $\{p_{k},\ket{\Psi_{k}}\}$ of the density matrix $\varrho$ we have
\begin{equation}\label{eq:boundavF}
\tfrac{1}{4}F_{Q}[\varrho,A]  \le  \sum_k p_k \va {A}_{\Psi_k} \le \va A_{\varrho},
\end{equation}
where the upper and the lower bounds are both tight in the sense that there are 
decompositions that saturate the first inequality, and there are others that saturate the second one \footnote{The fact that the second inequality of  \EQ{eq:boundavF}  can be saturated means that 
there is a decomposition such that the average variance of the pure states of the decomposition equals the variance of the mixed state.
It is possible to prove an analogous statement for $2\times2$ covariance matrices, while this is not always true for $3\times3$ covariance matrices  \cite{leka2013some, petz2013characterization}.}.

Let us now discuss an alternative way to interpret the inequalities of \EQ{eq:boundavF},
relating them to the theory of quantum purifications,
which play a fundamental role  in quantum information science.
A mixed state $\varrho$ with a decompostion \eqref{decomp} can be represented as a reduced state $\varrho={\rm Tr}_A(\ketbra{\Psi})$ 
of a pure state, called the purification of $\varrho,$ defined as 
\begin{equation}
\ket{\Psi}=\sum_k \sqrt{p_k} \ket{\Psi_k} \otimes \ket{k}_A.
\end{equation}
Here  $\ket{k}$ is an orthogonal basis for the ancillary system,
and ${\rm Tr}_A(.)$ denotes tracing out the ancilla.
Note that all purification can be obtained from each other using a unitary acting on the ancilla. 
This way one can obtain purifications corresponding to all the various decompositions.

Let us now assume that a friend controls the ancillary system and 
can assist us to achieve a high precision with the quantum state, or can
even hinder our efforts.
Our friend can choose between the purifications with unitaries acting on the ancilla.
Then, our friend makes a measurement on the ancilla in the  $\{\ket{k}\}$ basis
and sends us the result $k.$
This way we receive the states $\ket{\Psi_k}$ together with the label $k,$
corresponding to some decomposition of the type \eqref{decomp}.
The average 
quantum Fisher information for the $\ket{\Psi_k}$ states
 is bounded from below and from above as given in \EQ{eq:boundavF}.
The worst case bound is  given by the quantum Fisher information. We can always achieve this bound
 even if our friend acts against us.

On the other hand, 
if the friend acting on the ancilla helps us, 
a much larger average quantum Fisher information can be achieved, equal to four times
the variance. At this point, there is a further connection to quantum information science. Besides entanglement measures defined with convex roofs,
there are measures defined with concave roofs \cite{1998quant.ph..3033D,PhysRevLett.80.2493,2002quant.ph..6192L}. For example, the entanglement of assistance is defined as the maximum average entanglement that can be obtained if the party acting on the ancilla helps us.
Thus, in quantum information language, the variance can be called the quantum Fisher information of assistance over four.
Later, we will see another connection between purifications and the quantum Fisher information in \SEC{Sec:Metrology with noise}.

After the discussion relating the quantum Fisher information to the variance, and examining its convexity properties, we list some further useful
relations for the quantum Fisher information. From \EQ{eq:qF}, we can obtain directly the following identities.

(i) The formula \eqref{eq:qF} does not depend on the diagonal elements $\langle i \vert A \vert i \rangle.$ Hence,
\begin{equation}
F_Q[\varrho,A]=F_Q[\varrho,A+D],
\end{equation}
where $D$  is a matrix that is diagonal in the basis of the  eigenvectors of $\varrho,$ i.e., $[\varrho,D]=0.$

(ii) The following identity holds for all unitary dynamics $U$
\begin{equation}
\label{eq:SH}
F_Q[U \varrho U^\dagger ,A]=F_Q[\varrho,U^\dagger  A U].
\end{equation}
The left- and right-hand sides of \EQ{eq:SH} are similar to the Schr\"odinger picture and the Heisenberg picture, respectively,
in quantum mechanics.
Hence, in particular, the quantum Fisher information does not change under unitary dynamics governed by $A$ as a Hamiltonian
\begin{equation}
F_Q[\varrho,A]=F_Q[e^{-iA\theta} \varrho e^{iA\theta},A].
\end{equation}

(iii) The quantum Fisher information is additive under tensoring
\begin{equation}
F_{Q}[\varrho^{(1)}\otimes\varrho^{(2)},A^{(1)}\otimes\eins+\eins\otimes A^{(2)}]=F_{Q}[\varrho^{(1)},A^{(1)}]+F_{Q}[\varrho^{(2)},A^{(2)}].
\end{equation}
For $N$-fold tensor product of the system, we obtain an $N$-fold increase in the quantum Fisher information
\begin{equation}
F_{Q}[\varrho^{\otimes N},\sum_{n=1}^{N} A^{(n)} ]=NF_{Q}[\varrho,A],
\end{equation}
where $A^{(n)}$ denotes the operator $A$ acting on the $n^{\rm th}$ subsystem.

(iv) The quantum Fisher information is additive under a direct sum \cite{2014CoTPh..61...45L}
\begin{equation}\label{eq:directsum}
F_{Q}[\bigoplus_k p_k \varrho_k, \bigoplus_k A_k ]=\sum_k p_k F_{Q}[\varrho_k,A_k],
\end{equation}
where $\varrho_k$ are density matrices with a unit trace and $\sum_k p_k=1.$
\EQL{eq:directsum} is relevant, for example, for experiments where the particle number variance is not zero,
and the $\varrho_k$ correspond to density matrices with a fixed particle number \cite{PhysRevLett.105.120501,PhysRevA.86.012337}.

(v) If a pure quantum state $\ket{\Psi}$ of $N$ $d$-dimensional particles is mixed with white noise as \cite{PhysRevA.85.022321,PhysRevA.85.022322}
\begin{equation}
\varrho_{\rm noisy}(p) = p\ketbra{\Psi}+ (1-p)\frac{\openone}{d^N}, 
\end{equation}
then
\begin{equation}
F_{Q}[\varrho_{\rm noisy}(p),A]=\frac{p^2}{p+\frac{1-p}{2}d^{-N}}F_{Q}[\ketbra{\Psi},A].
\end{equation}
Thus, an additive global noise decreases the quantum Fisher information by a constant factor.
If $p$ does not depend on $N$ then it does not 
influence the scaling of the quantum Fisher information with the number of particles.
Note that this is not the case for a local uncorrelated noise. A constant uncorrelated local noise contribution can
destroy the scaling of the quantum Fisher information and lead back to the shot-noise scaling for large $N,$
as will be discussed in \SEC{Sec:Metrology with noise}.

(vi) If we have a bipartite density matrix and we trace out the second system,
the quantum Fisher information cannot increase (see, e.g. \REF{demkowicz2012elusive})
\begin{equation}\label{traceout}
F_Q[\varrho,A^{(1)} \otimes \openone^{(2)}] \ge F_Q [{\rm Tr_2}(\varrho),A^{(1)}].
\end{equation}
In fact, in many cases it decreases even if the operator $A^{(1)}$ was acting on the first subsystem,
and thus the unitary dynamics changed only the first subsystem. This is due to the fact that 
measurements on the entire system can lead to a better parameter estimation than measurements on
the first system. Let us see a simple example with the following characteristics
\bea
\varrho&=&\ketbra{\Psi^-},\nonumber\\
A^{(1)}&=&\sigma_z,
\eea
where \ket{\Psi^-} is defined in \EQ{singlet}. Since ${\rm Tr_2}(\varrho)$ is the completely mixed state,
the right-hand side of the inequality \eqref{traceout} is zero, while the left-hand side is positive.
On the other hand, \EQ{traceout} is always saturated if $\varrho$ is a product state of the form 
$\varrho^{(1)}\otimes\varrho^{(2)}.$

(vii) It is instructive to write the quantum Fisher information in an alternative form as \cite{PhysRevA.83.021804}
\begin{eqnarray}
F_{Q}[\varrho,A]&=&4\sum_{k,l}\lambda_{k}\vert \langle k \vert A \vert l \rangle \vert^{2}-8\sum_{k,l}\frac{\lambda_{k}\lambda_{l}}{\lambda_{k}+\lambda_{l}}\vert \langle k \vert A \vert l \rangle \vert^{2}\nonumber\\
&=&4\exs{A^2}-8\sum_{k,l}\frac{\lambda_{k}\lambda_{l}}{\lambda_{k}+\lambda_{l}}\vert \langle k \vert A \vert l \rangle \vert^{2}.
\label{eq:qF2}
\end{eqnarray}

(ix) Following a similar idea, \EQLL{convroof} can also be rewritten as 
\begin{equation}
F_{Q}[\varrho,A] =4\exs{A^2}_{\varrho}-4 \sup_{\{ p_k,\ket{\Psi_k}\}} \sum_k p_k \ex {A
 }^2_{\Psi_k}.\label{convroof_FQ}
\end{equation}
By removing the second moments of the operator from the infimum, 
we make the optimisation simpler.
Similarly, we can also rewrite the formula \EQ{convroof2} as\begin{equation}
\va{A}_{\varrho}=\exs{A^2}_{\varrho}-\inf_{\{ p_k,\ket{\Psi_k}\}} \sum_k p_k \ex {A
}^2_{\Psi_k}.\label{convroof222}
\end{equation}

(x) Finally, based on \EQ{convroof_FQ} and \EQ{convroof222}, the difference between the
variance and the quantum Fisher information over four is obtained as
\begin{equation}
\va{A}_{\varrho}-\frac{1}{4}F_{Q}[\varrho,A] =
\sup_{\{ p_k,\ket{\Psi_k}\}} \sum_k p_k \ex {A
 }^2_{\Psi_k}-
\inf_{\{ p_k,\ket{\Psi_k}\}} \sum_k p_k \ex {A
 }^2_{\Psi_k}.\label{convroof_FQQ}
\end{equation}
Clearly, \EQ{convroof_FQQ} is zero for all pure states. It can also be zero for some mixed states.
For example, based on \EQ{eq:Fvar},  we see that for all states for which we have $\va{A}_{\varrho}=0,$
we also have $F_{Q}[\varrho,A]=0.$ Thus, the difference \EQ{convroof_FQQ} is also zero for such quantum states.

\subsection{Optimal measurement}

The Cram\'er-Rao bound \eqref{eq:CR} defines the achievable largest precision of parameter estimation,
however, it is not clear what has to be measured to reach this precision bound.
An optimal measurement can be carried out if we measure in the eigenbasis of the
 symmetric logarithmic derivative $L$  \cite{PhysRevLett.72.3439,braunstein1996generalized}.
This operator is defined such that it can be used to
describe the quantum dynamics of the system with the equation
\begin{equation}\label{eq:LrrL}
\tfrac{d\varrho_\theta}{d\theta}=\tfrac{1}{2}(L\varrho_\theta+\varrho_\theta L).
\end{equation}
Unitary dynamics are generally given by the von Neumann equation with the Hamiltonian $A$
\begin{equation}\label{eq:LrrL2}
\tfrac{d\varrho_\theta}{d\theta}=i(\varrho_\theta A-A\varrho_\theta).
\end{equation}
The operator $L$ can be found based on knowing that the right-hand side of \EQ{eq:LrrL}
must be equal to the right hand-side of \EQ{eq:LrrL2}.
Hence, the symmetric logarithmic derivative can be expressed with a simple formula as 
\begin{equation}
\label{eq:L}
L=2i\sum_{k,l}\frac{\lambda_{k}-\lambda_{l}}{\lambda_{k}+\lambda_{l}} \vert k \rangle \langle l \vert \langle k \vert A \vert l \rangle,
\end{equation}
where $\lambda_k$ and $\vert k\rangle$ are the eigenvalues and eigenvectors, respectively, of the density matrix $\varrho.$
Based on \EQ{eq:FQ} and \EQ{eq:L}, the symmetric  logarithmic derivative  can be used to obtain the quantum Fisher information as
\begin{equation}
F_Q[\varrho,A]=\trace(\varrho L^2).
\end{equation}
For a pure state $\ket{\Psi},$ the formula \eqref{eq:L} can be simplified and the symmetric  logarithmic derivative can be obtained as
\begin{equation}
L=2i[\ketbra{\Psi},A].
\label{eq:qFpure}
\end{equation}

It is instructive to consider a concrete example. 
Let us find $L$ for the setup based on metrology with the fully polarised ensemble discussed in \SEC{Sec:Spin squeezed states}. In this case, $A=J_y$ and the quantum state
evolves according to the equation 
\begin{equation}
\varrho_\theta=e^{-iJ_y\theta} \varrho_0 e^{+i\theta J_y\theta},
\end{equation}
where the initial state is 
\begin{equation}
\varrho_0=\ketbra{0}^{\otimes N}.
\end{equation}
For short times, the dynamics can be written as 
\begin{equation}
\varrho_\theta\approx \varrho_0+ i\theta ( \varrho_0 J_y - J_y  \varrho_0).
\end{equation}
Using the identity with $2\times 2$ matrices
\begin{equation}\label{eq:xy}
i(\ketbra{0}j_y-j_y\ketbra{0})=\ketbra{0}j_x+j_x\ketbra{0}
\end{equation}
the short-time dynamics can be rewritten as
\begin{equation}
\varrho_\theta\approx\varrho_0+\theta ( \varrho_0 J_x + J_x  \varrho_0).
\end{equation}
Hence, for this case the symmetric logarithmic derivative is 
\begin{equation}
\label{eq:L1}
L=2J_x.
\end{equation}
Indeed, in the example of \SEC{Sec:Spin squeezed states} we measured $J_x,$
which now turned out to be the optimal operator to be measured.

Let us now see what can be obtained from the explicit formula \eqref{eq:qFpure} for the symmetric 
logarithmic derivative. Together with \EQ{eq:xy}, it leads to 
\begin{equation}
\label{eq:L2}
L=2(\ketbra{0}^{\otimes N} J_x+J_x\ketbra{0}^{\otimes N}).
\end{equation}
As the example shows,  \EQ{eq:L1}  and \EQ{eq:L2} are different, hence $L$ is not unique. 
Nevertheless, the right-hand side of \EQ{eq:LrrL} is the same for 
\EQ{eq:L1} and \EQ{eq:L2}. 
This is because the symmetric logarithmic derivative is defined 
unambiguously within the support of $\varrho_0,$ while in the orthogonal space
it can take any form as long as  \EQ{eq:LrrL}
is satisfied.

\subsection{Multi-parameter metrology}

The formalism of \SEC{Sec:Quantum Fisher information} can be generalized
to the case of estimating several parameters. The Cram\'er-Rao bound for this case is 
\begin{equation}\label{CR_mat}
C-F^{-1}\ge 0,
\end{equation}
where the inequality in \EQ{CR_mat} means that the left-hand side is a positive semidefinite matrix,
$C$ is now the covariance matrix with elements
\begin{equation}
C_{mn}=\exs{\theta_m \theta_n}-\exs{\theta_m}\exs{\theta_n},
\end{equation}
and $F$ is the Fisher matrix. It is defined as 
for the case of a unitary evolution
\begin{equation}
F_{mn}\equiv F_Q[\varrho,A_m,A_n]=
2\sum_{k,l}\frac{(\lambda_{k}-\lambda_{l})^{2}}{\lambda_{k}+\lambda_{l}} 
\langle k \vert A_m \vert l \rangle \langle l \vert A_n \vert k \rangle,\label{eq:qFmat}
\end{equation}
where $\lambda_k$ and $\ket{k}$ are the eigenvalues and eigenvectors of the density matrix $\varrho,$ respectively [see \EQ{decomp_eig}].

The bound of \EQ{CR_mat} cannot always be saturated, as it can happen
that the optimal measurement operators for the various $\theta_k$ parameters
do not commute with each other. Examples of multiparameter estimation 
include estimating parameters of unitary evolution as well as parameters of dissipative processes,
such as for example phase estimation in the presence of loss such that the loss is given \cite{PhysRevA.82.053804},
 the estimation of both the phase and the loss \cite{crowley2012multiparameter},
estimation of phase and diffusion in spin systems \cite{knysh2013estimation},
joint estimation of a phase shift and the amplitude of phase diffusion at the quantum limit \cite{vidrighin2014joint},
the joint estimation of the two defining parameters
of a displacement operation  (i.e., $x$ and $p$)  in phase space \cite{PhysRevA.87.012107},
optimal estimation of the damping constant and the reservoir temperature
\cite{PhysRevA.83.012315}, estimation of the 
 temperature and the chemical potential characterising quantum gases \cite{PhysRevA.88.063609}, 
estimation of two-parameter rotations in spin systems \cite {vaneph2013quantum},
and the simultaneous estimation of multiple phases \cite{PhysRevLett.111.070403}.
Multiparameter estimation is considered in a very general framework in \REF{0305-4470-35-13-307}.
Note that not all from the examples discussed above carry out a multi-parameter estimation in the sense it
was explained in this section.

\section{Quantum Fisher information and entanglement}
\label{sec:Quantum Fisher information and entanglement}

In this section, we review some important facts concerning the relation 
between the phase estimation sensitivity in linear interferometers and entanglement.
We will show that entanglement is needed to overcome the shot-noise
sensitivity in very general metrological tasks. Moreover, not only entanglement but 
multipartite entanglement is necessary for a maximal sensitivity. 
All these statements will be derived in a very general framework,
based on the quantum Fisher information.
We will also discuss related issues, namely, 
meaningful definitions of macroscopic entanglement,
the speed of the quantum evolution, and the quantum Zeno effect.
We will also briefly discuss the question whether inter-particle entanglement
is an appropriate notion for our systems.

\subsection{Entanglement criteria with the quantum Fisher information}
\label{sec:Entanglement_QF}

Let us first examine the upper bounds on the quantum Fisher information
 for general quantum states and 
for separable states.
These are also bounds for the sensitivity of the phase estimation, since 
due to the Cram\'er-Rao bound \eqref{eq:CR} we have
\be
(\Delta \theta)^{-2}\le F_Q[\varrho,J_l].
\ee

Entanglement has been recognised as an advantage for several metrological tasks (see, e.g., \REFS{sorensen2001many,PhysRevLett.100.100503}).
For a general relationship for linear interferometers, 
we can take advantage of the properties of the quantum Fisher
information discussed in \SEC{Sec:Quantum Fisher information}. Since for pure states the quantum Fisher
information equals four times the variance, for pure product states we can write
\begin{equation}\label{eq:fqbound}
F_{Q}[\varrho,J_{l}]=4\va{J_l}=4\sum_n \va{j_l^{(n)}}\le N
\end{equation}
for $l=x,y,z.$ For the second equality in \EQ{eq:fqbound},
we used the fact  that for a product state the variance of a collective observable
is the sum of the single-particle variances.
Due to the convexity of the quantum Fisher information, this upper bound is also valid
for separable states of the form \EQ{eq:sep} and we obtain \cite{PhysRevLett.102.100401} 
\begin{equation}
F_{Q}[\varrho,J_{l}]\le N.\label{eq:F2eb}
\end{equation}
All states violating \EQ{eq:F2eb}
are entangled. Such states make it possible to surpass the shot-noise limit and 
are more useful than separable states for some metrological tasks.

The maximum for general states, including entangled states, can be obtained similarly.
For pure states, we have
\begin{equation}
F_{Q}[\varrho,J_{l}]=4\va{J_l}\le N^2,\label{eq:He}
\end{equation}
which is a valid bound again for mixed states.
Thus, we obtained in \EQ{eq:F2eb} the shot-noise scaling \eqref{eq:shotnoise}, while in \EQ{eq:He} the Heisenberg scaling \eqref{eq:Heisenberg}
for the quantum Fisher information $F_{Q}[\varrho,J_{l}]. $
Note that our derivation is very simple,
and does not require any information about what we measure to estimate 
$\theta.$   Equation~\eqref{eq:F2eb} has already been used to detect entanglement based on
the metrological performance of the quantum states in \REFS{PhysRevLett.107.080504,lucke2011twin}.

At this point one might ask whether all entangled states can provide a 
sensitivity larger than the shot-noise sensitivity. This would show
that entanglement is equivalent to metrological usefulness.
Concerning linear interferometers,
it has been proven that not all entangled
states violate \EQ{eq:F2eb}, even allowing local unitary transformations. Thus, not all quantum states 
are useful for phase estimation
\cite{PhysRevA.82.012337}. 
It has been shown that there are even highly entangled pure states that are not useful.
Hence,
the presence of entanglement seems to be rather a necessary condition.

The quantum Fisher information can be used to define the entanglement parameter  \cite{PhysRevLett.102.100401}
\begin{equation}\label{eq:xi}
\chi^2=\frac{N}{F_{Q}[\varrho,J_{y}]}.
\end{equation}
Based on \EQ{eq:F2eb}, $\chi^2\ge1$ holds for separable states, while  $\chi^2<1$ indicates entanglement and 
also implies that the quantum state is more useful for metrology than separable states.
For pure states the new  parameter $\chi^2$ can be rewritten as
\begin{equation}\label{eq:xi2}
\chi^2=\frac{N}{4(\Delta J_{y})^2}.
\end{equation}
Thus, while $\xi_{\rm s}^2<1$ [$\xi_{\rm s}^2$ is given in \EQ{motherofallspinsqueezinginequalities}] indicates a small variance $(\Delta J_{x})^2,$
$\chi^2<1$ indicates a large variance $(\Delta J_{y})^2$ in the orthogonal direction.
Thus, the quantum state is metrologically useful not because it has a small variance 
along the $x$-direction, but because it has a large variance in the $y$-direction (see \FIG{fig:magneto}).

Next, we will relate the parameter 
\eqref{eq:xi} to the original spin squeezing parameter $\xi_s^2$ given in \EQ{motherofallspinsqueezinginequalities}.
\EQL{eq:accuracy} can be rewritten for the case that the axis of the rotation is in the $z-y$ plane, and it is not necessarily the $z$ axis as
\be\label{eq:zy}
(\Delta \theta)^2=\frac{\va{J_x}}{\exs{J_y}^2+\exs{J_z}^2}.
\ee
Combining \EQ{eq:zy} and the Cram\'er-Rao bound  \eqref{eq:CR} leads to \cite{PhysRevLett.102.100401}
\begin{equation}\label{eq:xichi}
\xi_{\rm s}^2\ge\chi^2.
\end{equation}
Hence, if $\xi_{\rm s}^2<1$ then  $\chi^2<1.$ 
Thus, the parameter $\chi^2$ is more sensitive to entanglement than $\xi_{\rm s}^2.$
One reason might be that $\xi_{\rm s}^2$ has information only about the reduced two-particle matrix of the state, as explained at the end of \SEC{sec:The original spin squeezing criterion}, while $\chi^2$ contains the quantum Fisher information that does not depend only on the two-particle state, but, in a sense, 
on the entire quantum state. In another context, we can say that due to \EQ{eq:xichi}, the spin squeezing parameter $\xi_{\rm s}^2$ detects entanglement that is
useful for metrology.

The previous ideas can be extended to construct
relations that include the quantum Fisher information corresponding to several metrological tasks. In order to construct such a relation,
let us consider the average quantum Fisher information
for any direction defined as
\begin{equation}
\label{eq:avg}
{\rm avg_{\vec n}} F_Q[\varrho,J_{\vec n}]=\int_{\vert \vec n\vert=1} F_Q[\varrho,J_{\vec n}] d{\vec n}.
\end{equation}
\EQL{eq:avg}  is relevant for the following metrological task.
It gives an upper bound on the average $(\Delta\theta)^{-2}$ for a quantum state $\varrho,$ if the direction of the magnetic field
is chosen randomly based on a uniform distribution.

Simple calculations show that the integral \eqref{eq:avg} equals the average of the 
quantum Fisher information corresponding to the three angular momentum components
\begin{equation}\label{eq:av3}
{\rm avg_{\vec n}} F_Q[\varrho,J_{\vec n}]=\tfrac{1}{3}(F_Q[\varrho,J_x]+F_Q[\varrho,J_y]+F_Q[\varrho,J_z]).
\end{equation}
Bounds similar to \EQ{eq:F2eb} can be obtained also for 
separable states for the 
average quantum Fisher information
\cite{PhysRevA.85.022321,PhysRevA.85.022322}.
 It can be proven that for separable states 
\begin{equation}\label{eq:sepmax}
{\rm avg_{\vec n}} F_Q[\varrho,J_{\vec n}] \le \tfrac{2}{3} N 
\end{equation}
holds. Comparing \EQ{eq:sepmax} with \EQ{eq:F2eb} shows that the bound for the average
quantum Fisher information is lower
than the bound for quantum Fisher information for a single metrological task for a given direction. The bound for all quantum states, including entangled states is
\begin{equation} \label{eq:avgmax}
{\rm avg_{\vec n}} F_Q[\varrho,J_{\vec n}] \le \tfrac{1}{3} N(N+2). 
\end{equation}
The bound for the average quantum Fisher information is again smaller than the bound for 
a given direction appearing in \EQ{eq:He}.

Let us now calculate the quantum Fisher information for
 concrete highly entangled quantum states.
For GHZ states \eqref{eq:GHZ}, the quantum Fisher information values 
for the three angular momentum components are
\be\label{eq:fqghz}
F_Q[\varrho,J_x]=N^2,\;\;\;\;\;F_Q[\varrho,J_y]=F_Q[\varrho,J_z]=N,\nonumber\\
\ee
while for $N$-qubit
symmetric Dicke states with $\tfrac{N}{2}$  excitations given in \EQ{eq:Dicke}
we have 
\be\label{eq:fqdicke}
F_Q[\varrho,J_x]=F_Q[\varrho,J_y]=\tfrac{1}{2}N(N+2),\;\;\;\;\;F_Q[\varrho,J_z]=0.\nonumber\\
\ee
Hence, GHZ states have a maximal sensitivity for the metrology of the type considered in
\SEC{sec:Metrology with a GHZ state} and hence saturate \EQ{eq:He}. 
As can be seen in \EQ{eq:fqdicke}, Dicke states \eqref{eq:Dicke} almost reach the maximum for $F_Q[\varrho,J_x]$ and $F_Q[\varrho,J_y].$ 
Note that both states are unpolarised, that its, their mean spin is zero. It is easy to show that $F_Q[\varrho,J_l]$ can be maximal
only if $\exs{J_l}=0,$ which can be seen from the inequality 
\be\label{eq:F_meanspin1}
F_Q[\varrho,J_l]\le 4\va{J_l}\le N^2-4\exs{J_l}^2.\nonumber\\
\ee

Let us turn to the average sensitivity of the quantum states.
Both GHZ states \eqref{eq:GHZ} and 
symmetric Dicke states \eqref{eq:Dicke} saturate the inequality for the average quantum Fisher information~\eqref{eq:avgmax}.
In general, simple algebra shows that 
all pure symmetric states for which $\langle J_{l}\rangle=0$
for $l=x,y,z$ saturate \EQ{eq:avgmax}, that is, their average sensitivity is maximal \cite{PhysRevA.85.022322}.
This indicates that states without a large spin can be more useful for metrological purposes than polarised quantum states. 

Let us formulate this statement in a more quantitative way, by bounding the average 
quantum Fisher information 
with the spin length. We can construct such  an inequality using
\eqref{varlength} and \eqref{eq:Fvar} as
\be\label{Flimit}
F_Q[\varrho,J_x]+F_Q[\varrho,J_y]+F_Q[\varrho,J_y]\le 4 (\exs{J_x^2+J_y^2+J_z^2}-\exs{\vec J}^2),
\ee
where the mean spin vector is defined as
\be
\exs{\vec J}=(\exs{J_x},\exs{J_y},\exs{J_z}).
\ee
Equation \eqref{Flimit} expresses the fact that a large average precision \eqref{eq:av3}
can be reached if the state is close to symmetric and thus the inequality  \eqref{Jxyz2} is close to being saturated.
Moreover, for a large average precision, 
$\vert \exs{\vec J} \vert$ must be small. The maximal average 
precision can be reached only if the state is symmetric and 
 $\exs{\vec J}=0.$
Hence, states that are almost fully polarised have an average precision that is far from the maximum. 
With this we generalised the discussions at the end of \SEC{Sec:Spin squeezed states}.

\subsection{Criteria for multipartite entanglement}
\label{sec:Multipartite entanglement}

After defining the basic notions, we will find the bounds
for the metrological sensitivity of quantum states with
various levels of multipartite entanglement.
For $N$-qubit $k$-producible states, 
the quantum Fisher information is bounded from above by  \cite{PhysRevA.85.022321,PhysRevA.85.022322}
\begin{equation}
F_{Q}[\varrho,J_{l}]\le sk^2+(N-sk)^2,\label{eq:Nka-2}
\end{equation}
where $s$ is the integer part of $\tfrac{N}{k}.$ It is instructive 
to write \EQ{eq:Nka-2} for the case $N$ divisible by $k$ as
\begin{equation}\label{eq:Nkentangl}
F_Q[\varrho,J_{\vec n}]  = Nk.
\end{equation}
Thus, the bounds reached by $k$-producible states are distributed linearly, i.e., $(2k)$-poducible states
can reach a twice as large value for $(\Delta \theta)^{-2}$ as $k$-producible states can.

Similar bounds can be obtained for the average 
quantum Fisher information.
 For $N$-qubit $k$-producible
states, for $k\ge2,$ the sum of the three Fisher information terms is bounded from
above by \cite{PhysRevA.85.022321,PhysRevA.85.022322}
\begin{eqnarray}
{\rm avg_{\vec n}} F_Q[\varrho,J_{\vec n}]\le
\Bigg\{
\begin{array}{l}
\tfrac{1}{3}sk(k+2)+\tfrac{1}{3}(N-sk)(N-sk+2)\nonumber\\
\tfrac{1}{3}sk(k+2)+\tfrac{2}{3}
\end{array} & \begin{array}{l}
\textrm{if }N-sk\ne1,\\
\textrm{if }N-sk=1,
\end{array}\\
\label{eq:Nka}
\end{eqnarray}
where $s$ is again the integer part of $\tfrac{N}{k}.$ Any state that
violates this bound is not $k$-producible and contains $(k+1)$-particle
entanglement. These inequalities have been used to detect experimentally useful multipartite entanglement in \REF{PhysRevLett.107.080504}.

It is also instructive to find bounds for genuine
$N$-particle entanglement [see \SEC{sec:Entanglement and multi-particle entanglement}]. The bounds for biseparable states for the left-hand side
of \EQ{eq:Nka-2} and \EQ{eq:Nka} can be obtained
taking $n=1$ and maximizing the bounds over $k=\frac{N}{2},\frac{N}{2}+1,...,N-1$ for even $N,$ while
over $k=\frac{N+1}{2},\frac{N+1}{2}+1,...,N-1$ for odd $N.$
Hence, we arrive at
\begin{subequations}
\begin{eqnarray}
F_{Q}[\varrho,J_{l}] &\le& (N-1)^{2}+1,\label{eq:bisep-1}\\
{\rm avg_{\vec n}} F_Q[\varrho,J_{\vec n}]&\le& \tfrac{1}{3}(N^{2}+1).\label{eq:bisep}
\end{eqnarray}
\end{subequations} 
Any state that violates \EQ{eq:bisep-1}
or \EQ{eq:bisep} is genuine multipartite entangled. 
Comparing these to the bounds for general entangled states,
\EQS{eq:He} and \eqref{eq:avgmax},
we can conclude that full $N$-partite entanglement is needed to 
reach a maximal metrological sensitivity. 

Finally, we also mention that bound entangled states can also be detected 
with the entanglement criteria based on the quantum Fisher information.
Bound entanglement is a weak type of entanglement,
which is not distillable with local operations
and classical communication \cite{RevModPhys.81.865,guhne2009entanglement}.
Thus, it is a surprise that such states are useful for metrology.
\REFL{PhysRevA.85.022321} presented states that were detected as bound entangled based on the
criterion for the average quantum Fisher information \eqref{eq:sepmax}.
\REFL{2014arXiv1403.5867C} presented states that violate 
the criterion based on a bound for the quantum Fisher information for a single metrological task \eqref{eq:Nka-2}
and showed that even the Heisenberg scaling can be reached with
bound entangled states. 

\begin{figure}
\begin{center}
\includegraphics[width=6cm]{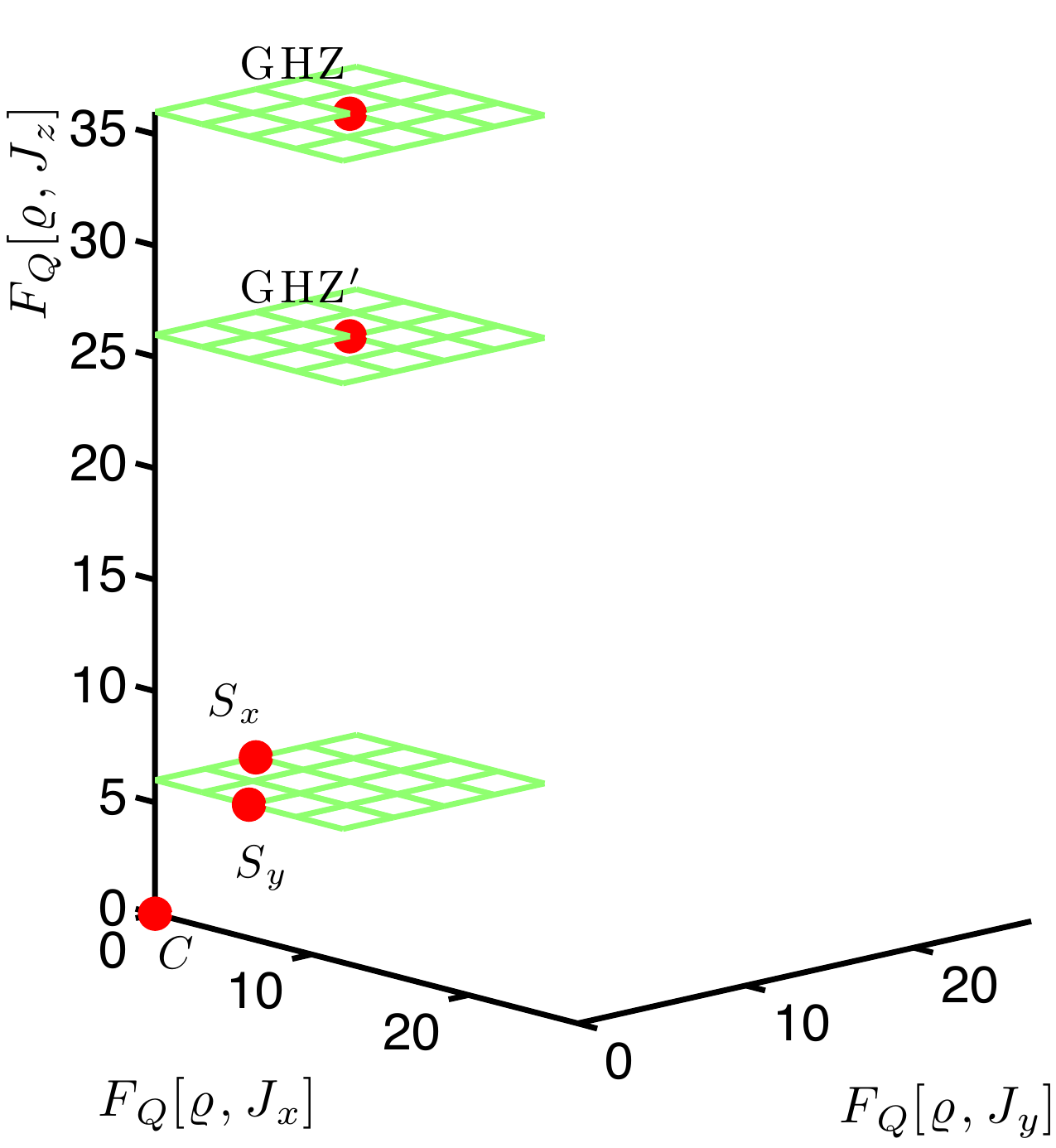}
\includegraphics[width=6cm]{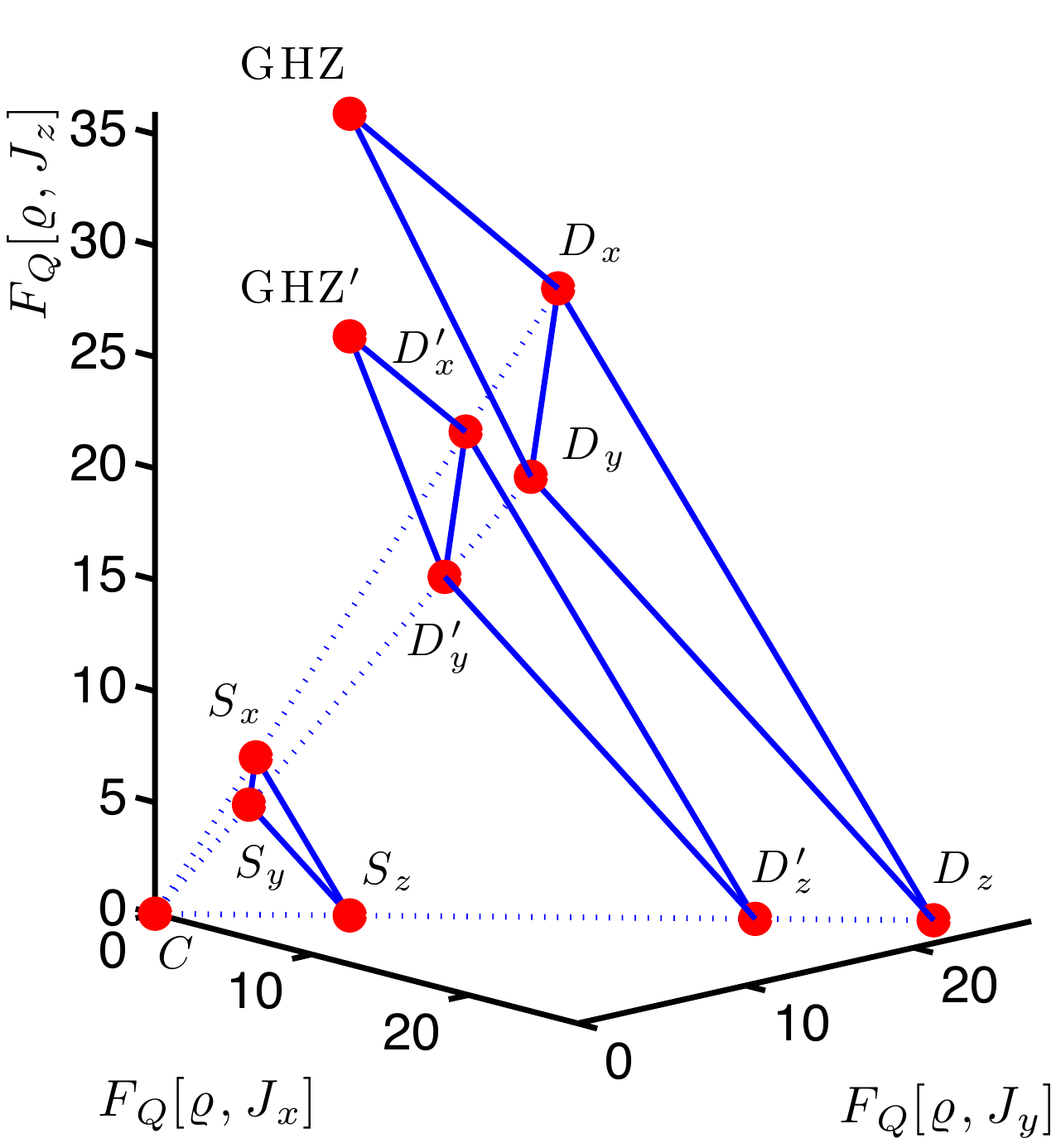}

(a) {\hskip 4cm} (b)
\caption{
Entanglement criteria in the $(F_{Q}[\varrho,J_x], F_{Q}[\varrho,J_y], F_{Q}[\varrho,J_z])-$space for $N=6$ qubits. 
(a) Entanglement criteria
with $F_{Q}[\varrho,J_z].$ The planes correspond to the inequalities, from bottom to top, Eqs.~\eqref{eq:F2eb}, \eqref{eq:bisep-1} and \eqref{eq:He} with $J_l=J_z.$ (b) 
Entanglement criteria with $\sum_l F_{Q}[\varrho,J_l].$ The planes correspond to the  inequalities, from bottom to top,  Eqs.~\eqref{eq:sepmax}, \eqref{eq:bisep} and \eqref{eq:avgmax}, respectively. For the description of the inequalities and the points see text.}
\label{fig:Fxyz}
\end{center}
\end{figure}

The results presented in this section are summarized in \FIG{fig:Fxyz}, which shows various quantum states in the  $(F_{Q}[\varrho,J_x], F_{Q}[\varrho,J_y], F_{Q}[\varrho,J_z])-$space for $N=6$ particles. A similar figure can be drawn for any even $N.$

In \FIG{fig:Fxyz}(a), the entanglement conditions given in terms of an inequality with the quantum Fisher information $F_{Q}[\varrho,J_z]$ are shown.  The completely mixed state corresponds to the point $C$ at the origin.
The entanglement criterion~\eqref{eq:F2eb} corresponds to the bottom plane. Any state above this plane is entangled. For even $N,$ product states corresponding to the $S_l$ points are defined as
\be\label{eq:Sl}
\varrho_{S_l}=\left\vert 0 \right\rangle_l\left\langle 0 \right\vert ^{\otimes \frac{N}{2}}\otimes 
\left\vert 1 \right\rangle_l\left\langle 1\right\vert ^{\otimes \frac{N}{2}}
\ee
for $l=x,y,z.$ The entanglement criterion~\eqref{eq:bisep-1} corresponds to the second plane. Any state above this plane is genuine multipartite entangled. The point GHZ' refers to the state
\be
\varrho_{{\rm GHZ}'}=\ketbra{{\rm GHZ}_{N-1}}\otimes \frac{\openone}{2}.
\ee
Finally, the inequality~\eqref{eq:He} corresponds to the top plane. The point GHZ in \FIG{fig:Fxyz}(a) refers to the GHZ state  given in \EQ{eq:GHZ}.

Let us now turn to \FIG{fig:Fxyz}(b), in which the entanglement conditions involving the average 
quantum Fisher information \eqref{eq:av3}
are shown. The plane of the triangle $S_x-S_y-S_z$ corresponds to the inequality \eqref{eq:sepmax},
all states  
 above this plane are entangled.
States corresponding to the points $S_l$ are given in \EQ{eq:Sl}.
The ${\rm GHZ}'-D_x'-D_y'-D_z'$ plane corresponds to  the inequality \eqref{eq:bisep}, and
all states  
above this plane are genuine multipartite entangled.
A tensor product  of a Dicke state and  a single-particle state corresponds to the point $D_z'$
\be
\ket{D_z'}=\left\vert D_{(N-1)}^{(\frac{N}{2}-1)}\right\rangle \otimes \left\vert 1 \right\rangle,
\ee
where we used the definition of the symmetric Dicke state given in \EQ{eq:DickeNm}. 
Note that since $(N-1)$ is odd, $\left\vert D_{(N-1)}^{(\frac{N}{2}-1)}\right\rangle$ has unequal number of $0$'s and $1$'s.
The points $D_x'$ and $D_y'$ are defined similarly with Dicke states in the $x$ and $y$-basis, respectively.
The ${\rm GHZ}-D_x-D_y-D_z$ plane corresponds to  
 \EQ{eq:avgmax}.
There are no physical states 
above this plane.
The Dicke state in the $z$-basis characterised by \EQ{eq:fqdicke} corresponds to the point $D_z.$
The Dicke states in the $x$ and $y$-basis correspond to the points $D_x$ and $D_y,$ respectively.
GHZ states given in the $x$ and $y$-basis, not shown in the \FIG{fig:Fxyz}(b), would also be in the same plane. 

\subsection{Macroscopic superposition}
\label{Sec:Macroscopic superposition}

After describing the relation between the quantum Fisher information and multiparticle entanglement, we will discuss one of the related interesting questions, namely, the proper definition of a macroscopic superposition  
\cite{PhysRevLett.95.090401,1367-2630-14-9-093039,PhysRevLett.106.110402}. 
The question whether a quantum state is a macroscopic superposition makes sense
only if we consider a state that can be defined for any number of particles, such as the
GHZ state. Then, we examine the properties of this family of states for an increasing $N.$ 

A very meaningful definition has been proposed for pure states based on the variance of collective quantities as follows \cite{PhysRevLett.95.090401}. Let us define ${\mathcal {A}_{\rm coll}}$ as the set of collective operators 
given as
\be\label{Aop}
A_{\rm coll}=
\sum_{n=1}^N a^{(n)}.
\ee
Here $a^{(n)}$ are single-particle operators acting on the $n^{\rm th}$ particle and for the operator norm we require 
$\vert\vert a^{(n)}\vert\vert =1.$ Note that before $A$ was just a general Hermitian operator, however, from now on 
we require that it is the sum of single-particle operators. For spin-$\frac{1}{2}$ particles and traceless $a^{(n)},$ the operators of the type \eqref{Aop} are just the collective angular momentum component $J_z$ apart from local unitaries and a constant factor.

Then, we can define the index $p$ as
\be\label{eq:index}
\max_{ A \in{\mathcal {A}_{\rm coll}}} (\Delta A
)^2=\mathcal{O}(N^p).
\ee
Here, $f(x)=\mathcal{O}(x^m)$ means that $\lim_{x\rightarrow\infty} \frac{f(x)}{x^m}={\rm constant} >0.$
The index $p$  in \EQ{eq:index} is confined in a range $1\le p \le 2$ and for any product state we have $p=1. $
The state is called macroscopically entangled if $p=2.$
These ideas can be extended to mixed states using the quantum Fisher information rather than the variance \cite{1367-2630-14-9-093039,PhysRevLett.106.110402}.
The effective size of a macroscopic superposition is defined as
\be
N_{\rm eff}^F(\varrho)=\frac{1}{4N}\max_{A
\in{\mathcal {A}_{\rm coll}}} F_Q[\varrho,A
].
\ee
We call a quantum state a macroscopic superposition if $N_{\rm eff}^F(\varrho) = \mathcal{O}(N).$ 

We can compare this definition to our findings concerning the relationship between
 multipartite entanglement and 
the quantum Fisher information in \SEC{sec:Multipartite entanglement}. Based on \EQ{eq:Nkentangl}, we can see that for $N$ divisible by $k,$ for $k$-producible states we have $N_{\rm eff}^F(\varrho) = k.$
Hence, for a quantum state to be a macroscopic superposition,  it is necessary that the state is $k$-particle entangled  such that $k=\mathcal{O}(N).$
Thus, $k$ must grow linearly with $N,$ while we do not need to have full $N$-body entanglement.

Since the quantum Fisher information is convex, as discussed in 
\SEC{Sec:Quantum Fisher information}, if we mix two states with each other that are not macroscopic quantum superpositions, we will always get
 a state that is  not a macroscopic quantum superposition either.
Hence, states that are not  macroscopic quantum superpositions form a convex set.
This is, of course, expected for a meaningful definition.
Moreover, since the quantum Fisher information is the convex roof of the variance, as discussed in \SEC{Sec:Quantum Fisher information}, it is always
larger or equal to other convex functions that for pure states equal four times the variance.
Hence our definition 
gives a larger set of 
states forming a macroscopic superpositions than if we used instead of 
the quantum Fisher information \eqref{eq:qF}
generalised quantum Fisher informations,
 like four times the Wigner-Yanase skew information \eqref{eq:WY} mentioned earlier.

\subsection{Speed of quantum evolution}
\label{Sec:Speed of quantum evolution}

In this section, we discuss that 
the entanglement properties of a quantum state are related to the speed of the quantum evolution starting out from the state, via the quantum Fisher information \cite{PhysRevLett.110.050403,PhysRevLett.110.050402}. We will show that states with a 
large quantum Fisher information, and thus, a large multiparticle entanglement can evolve faster than states with a small quantum Fisher information.

We use the Bures fidelity to describe the speed of the evolution.
This quantity plays a central role in quantum information science.
It is defined as
\be
F_B(\varrho_1,\varrho_2)=
{\rm Tr}(\sqrt{\sqrt{\varrho_1}\varrho_2\sqrt{\varrho_1}})^2.
\ee
Clearly, $0\le F_B(\varrho_1,\varrho_2)\le 1.$
The fidelity is $1$ only if $\varrho_1=\varrho_2,$ while it is $0$ if the two states live in orthogonal subspaces. 
If $\varrho_1$ is a pure state, then the fidelity can be expressed in a simpler form
\be
\label{eq:FB2}
F_B(\varrho_1,\varrho_2)=
{\rm Tr}(\varrho_1\varrho_2).
\ee
If both states are pure, then the fidelity is just the square of the absolute value of the overlap
\be
\label{eq:FB2b}
F_B(\ket{\Psi_1},\ket{\Psi_2})=
\vert\langle \Psi_1\vert\Psi_2\rangle\vert^2.
\ee

The fidelity can be used to characterise the speed of the evolution by
calculating the fidelity between the initial state $\varrho$ and the final state $\varrho_{\theta}.$ 
Using the fact that the quantum Fisher information
is proportional to the second derivative of the fidelity with respect to ${\theta}$
in this case, we can write \cite{PhysRevLett.110.050403,PhysRevLett.110.050402}
\be\label{eq:FB}
F_B(\varrho,\varrho_{\theta})=
1-\theta^2\tfrac{F_Q[\varrho,A]}{4}+\mathcal{O}(\theta^3),
\ee
where the parameter $\theta$ is small and the
system Hamiltonian is given by  $A. $ Note that there cannot be a term linear in $\theta$ since such a term
would result in $F_B(\varrho,\varrho_{\theta})>1$ either for $\theta<0$ or for $\theta>0$ for small $\vert \theta \vert.$
Apart from the expansion \eqref{eq:FB}, a bound for the fidelity can be obtained from 
the improved Madelstam-Tamm bound \cite{PhysRevA.85.052127}
\be\label{eq:frowis}
F_B (\varrho,\varrho_{\theta}) \ge \cos^2 \left(\sqrt{\tfrac{F_Q[\varrho,A]}{4}}\theta\right)
\ee
with the condition
\be
\sqrt{F_Q[\varrho,A]} \vert \theta \vert \le \pi.
\ee
Based on the properties of the fidelity described above, 
we can interpret the relation \eqref{eq:FB} in the following way.
As $\theta$\ is increasing, the fidelity with respect to the initial state
starts to decrease.
The quantum Fisher information $F_Q[\varrho,A]$ in  \EQ{eq:FB} 
characterises the speed at which the quantum state
evolves into a state orthogonal to the initial state. 

If $A$ is the sum of single-particle operators,  i.e., $A\in{\mathcal {A}_{\rm coll}}$  then
the speed of the evolution is related to the entanglement of the quantum state.
One can see that large quantum Fisher information is needed for a large speed. Based on \SEC{sec:Multipartite entanglement}, it is also clear that for a large quantum Fisher information a large multipartite entanglement is needed. Thus, large multiparticle entanglement is needed
for a large speed of the quantum evolution.

\subsection{Quantum Zeno effect}

The quantum Zeno effect is one of the most discussed paradoxical 
features of quantum mechanics. It has been recently shown that it is related to the 
quantum Fisher information.  Due to the Zeno effect,
if we have a quantum system in an initial state, that starts to evolve under a Hamiltonian,
and we perform a projective measurement, projecting the state to the initial state, with a sufficient frequency then the system will
be unable to evolve, and stays in the initial state. This can be expressed more quantitatively as follows.
If we perform $m$ measurements at times $t_k=k\tau$ for $k=0,1,..,m$ such that $m\rightarrow \infty,$
$\tau\rightarrow 0,$ and $m\tau=t,$ then we find that the quantum system does not evolve and it stays in its initial state.
It turns out that the characteristic quantum Zeno time is \cite{PhysRevLett.109.150410, schafer2014experimental}
\be\label{eq:Zeno}
\tau_{QZ}=\frac{2}{\sqrt{F_Q[\varrho,A]}}.
\ee
In order to see the quantum Zeno effect, we need to perform the projective measurements with a frequency larger than $1/\tau_{QZ}.$

As in the case of the speed of the quantum evolution, it is interesting to consider the case when
$A$ is the sum of single-particle operators. 
Then, based on \EQ{eq:Zeno}, we see that for states with a large quantum Fisher information $\tau_{QZ}$ is small.
Hence, based on \SEC{sec:Multipartite entanglement}, for entangled states $\tau_{QZ}$ can be $\sim \frac{1}{N}$ while for separable states we obtain $\tau_{QZ}\sim  \frac{1}{\sqrt{N}}.$ These show that the entanglement properties of a quantum state are reflected in its behaviour in the quantum Zeno effect.

\subsection{Multi-particle entanglement vs. mode-entanglement }
\label{Sec:Multi-particle entanglement vs. mode-entanglement}

In this section, we discuss the meaning of multipartite entanglement of very many particles. 
The full discussion of this topic is outside of the scope of our review. However, we would mention
connections of this problem to the topics covered here.

Entanglement is typically considered between two- or more parties that are spatially separated from each other and are individually accessible. This is due to the fact that entanglement theory developed from the theory of Bell inequalities that required even a space-like separation of the detection events at the parties \cite{RevModPhys.81.865,guhne2009entanglement}.
Moreover, distillation of entanglement by local operations and classical communication, and many other quantum information processing tasks need a local access to the particles \cite{RevModPhys.81.865,guhne2009entanglement}.

However, in a many-particle system, where we have $10^6-10^{12}$ particles, this picture cannot be 
maintained, since even with a large technological advancement we would not 
be able to access the particles individually. The situation is even more complicated when
we consider Bose-Einstein condensates of two-state atoms. In this case, ideally, all the atoms 
occupy the same spatial state. 

Internal quantum states of Bose-Einstein condensates of two-state atoms can be written as an $N$-qubit symmetric quantum state or as a two-mode quantum state. The two descriptions are equivalent  to each other. For example, the GHZ state as a multiparticle state is given in \EQ{eq:GHZ}, while as a two-mode state can be given in \EQ{eq:NOON}. It can happen that a state that is highly entangled in one description is not entangled in the other, and vice versa. 
It has been argued by some authors that for states of Bose-Einstein condensates, the entanglement between the modes has to be considered rather than entanglement between particles \cite{PhysRevA.87.012340,Benatti2010924}. 
It has been shown that the precision can surpass the shot-noise limit 
 in a linear interferometer with non-entangled states, i.e., 
 states without entanglement between the modes \cite{PhysRevA.87.012340}.

The previous sections of this review help us to analyse this question from the point of view of quantum metrology.
As discussed in \SEC{sec:Entanglement_QF}, inter-particle entanglement is needed to overcome the shot-noise limit for linear interferometers. The entanglement condition based on metrological usefulness 
is equally valid for ensembles of individually accessible particles and ensembles of particles that are not accessible individually. In fact, it is also valid for quantum states of Bose-Einstein condensates.  If the state satisfies the definition of separability \eqref{eq:sep} then it cannot surpass the shot-noise limit, even in a Bose-Einstein condensate. The same is true for multiparticle entanglement, as discussed in \SEC{sec:Multipartite entanglement}. Full multiparticle entanglement is needed to reach the maximum precision. 
Moreover, the definition of macroscopic superposition given in \SEC{Sec:Macroscopic superposition} does not require individually accessible particles.
Finally, this is also true for our findings relating the quantum Fisher information to the speed of evolution discussed in \SEC{Sec:Speed of quantum evolution}.
There is a maximal speed for separable multi-particle states satisfying \eqref{eq:sep}, regardless of whether the particles can be individually accessed or not. 
Based on these, it seems to be reasonable to say that inter-particle entanglement is a very useful notion even for large ensembles and even for a Bose-Einstein condensate of two-state atoms. 

\begin{figure}
\begin{center}
\includegraphics[width=7cm]{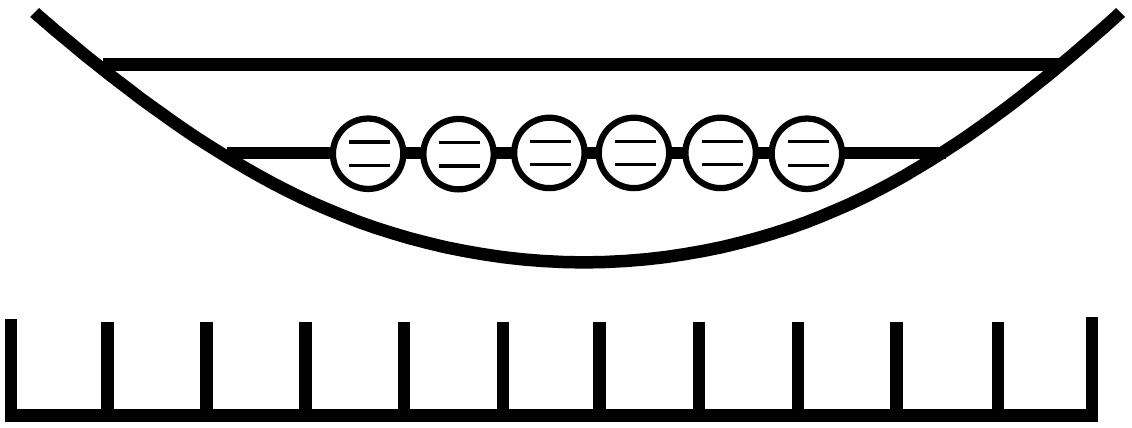} 
\caption{
The dilute cloud argument. While in the Bose-Einstein condensate, the two-state particles occupy the same spatial mode. If we let the could expand, and we detect the particle positions with detectors that do not disturb the internal state of the atoms, we obtain an ensemble of distinguishable particles with the same internal state that they had in the Bose-Einstein condensate. If we have much more detectors than particles in a dilute cloud, we can achieve that no two particles end up at the same detector.}
\label{fig:cloud}
\end{center}
\end{figure}

Still, it remains an important question whether the interparticle entanglement within Bose-Einstein condensates can be converted into entanglement between individually accessible particles.
An insight concerning this question can be gained from the dilute cloud argument \cite{PhysRevA.86.012337}. This thought-experiment, shown in \FIG{fig:cloud}, demonstrates that an ensemble of two-state particles in a Bose-Einstein condensate can be converted to an ensemble of localised particles, while the internal state of the ensemble remains unchanged. Finally, entanglement within a Bose-Einstein condensate of two-state atoms can be converted to entanglement between two spatial modes of the same particles  with straightforward non-entangling operations \cite{PhysRevLett.112.150501}. This statement can even be generalised to the case of splitting the condensate into three or more ensembles entangled with each other.

\section{Metrology in a noisy environment}
\label{Sec:Metrology with noise}

In this section, we examine what happens if noise appears in the quantum metrological setup. 
We will show that uncorrelated
 noise can easily destroy the Heisenberg scaling \eqref{eq:Heisenberg} and restore 
the shot-noise scaling \eqref{eq:shotnoise}, above a certain particle number.

In order to gain an intuitive understanding, let us consider the example of magnetometry with spin squeezed states discussed in \SEC{Sec:Spin squeezed states} and examine the effect of uncorrelated noise.
Let us assume that a particle with a state $\varrho_1$ passes trough a map that turns its internal state
 to the fully mixed state with some probability $p$ as
\be
\epsilon_p(\varrho_1)=(1-p)\varrho_1+p\tfrac{\openone}{2}.
\ee
If this map acts in parallel on all particles, the state can be given as a mixture
\be\label{epsrho}
\epsilon_p^{\otimes N}(\varrho)=\sum_{n=0}^N p_n \varrho_n,
\ee
where the state  obtained after $n$ particles decohered into the completely mixed state is 
\be
\varrho_n= \tfrac{1}{N!}\sum_k \Pi_k\left[\left(\tfrac{\openone}{2}\right)^{\otimes n}\otimes{\rm Tr}_{1,2,...,n}(\varrho)\right]\Pi_k^\dagger.
\ee
Here the summation is over all permutations $\Pi_k$, since
we took advantage of the fact that the spin squeezed state is permutationally invariant.
This is a reasonable assumption, as both the initial state and the spin squeezing dynamics are typically permutationally invariant.
The  quantity $p_n$ in the decomposition \eqref{epsrho} is obtained with the binomial coefficients as
\be\label{eq:pn}
p_n=\binom{N}{n}p^n(1-p)^{(N-n)}.
\ee

For the noisy state \eqref{epsrho}, the variance of the collective angular momentum component can be bounded from below as
\be\label{epsrho3}
\va{J_x}\ge \sum_n p_n \va{J_x}_{\varrho_n} \ge  \sum_n p_n \tfrac{n}{4} = \tfrac{pN}{4}.
\ee
For the first inequality in \EQ{epsrho3}, we used the concavity of the variance.
For the second   inequality, we used the fact  that for a product state of the form
${\varrho}={\varrho_A\otimes \varrho_B},$ the variance is the sum of the variances for $\varrho_k$ 
\be
\va{J_x}_{{\varrho}}=\va{J_x}_{{\varrho_A}}+\va{J_x}_{{\varrho_B}}.
\ee
Note that $\varrho_A$ and  $\varrho_B$ 
are possibly multi-particle states.
To obtain the bound on the right-hand side of \EQ{epsrho3}, we used a well-known identity
for the binomial coefficients
\be\label{eq:bin}\sum_n n p_n=pN,\ee
where $p_n$ is defined in \EQ{eq:pn}.
The equality \eqref{eq:bin} expresses the fact 
that the expectation value of the number 
of spins that undergo decoherence is $pN.$
In the bound on the variance \eqref{epsrho3} the term linear in $N$ appears since, in average, the
 same fraction of the spins is affected for all $N$ and all the decohered spins 
introduce the same additional noise for the collective angular momentum component.
Based on these and on the formula \eqref{eq:accuracy} giving the precision,
we obtain 
\be\label{epsrho2}
(\Delta \theta)^{-2} \le \frac{\frac{N^2}{4}}{\frac{pN}{4}}  \propto N,
\ee
which means that the precision has a shot-noise scaling for large $N.$ It is important to note that
this is true whenever $p>0,$ hence any nonzero decoherence will lead to a shot-noise scaling.

After this instructive example, let us review the existing literature on the subject.
The effect of the noise has been analysed for a setup based on metrology with GHZ states described in 
\SEC{sec:Metrology with a GHZ state}. Let us consider the dynamics
\be
\frac{d\varrho}{dt}=-i[\gamma B J_z,\varrho]+\epsilon_{xyz}^{\otimes N}(\varrho),
\ee
where the first term describes the coherent evolution, while the second term is the decoherence.
The single-particle decoherence is given as
\be\label{eq:LK}
\epsilon_{xyz}(\varrho_1)= -\tfrac{\gamma_{\rm n}}{2} \left( \varrho_1-\alpha_x \sigma_x \varrho_1 \sigma_x -\alpha_y \sigma_y \varrho_1 \sigma_y-\alpha_z \sigma_z\varrho_1 \sigma_z \right),
\ee
where $\gamma_{\rm n}$ is the overall noise strength, $\varrho_1$ is a single-particle state, $\sigma_l$ are the Pauli spin matrices,  $\alpha_l\ge 0,$ and  $\alpha_x+\alpha_y+\alpha_z=1.$ 
In \REF{PhysRevLett.79.3865}, it has been shown that for $\alpha_z=1$ the GHZ state does not provide a better scaling of the precision with the particle number than uncorrelated states. On the other hand, states leading to an optimal precision are different from the GHZ state. 

Recently, it has been proven in a very general framework that in the presence of even very small decoherence, $(\Delta \theta)^{-2}$ scales with $N$ rather than with $N^2$ for large particle numbers \cite{escher2011general,demkowicz2012elusive,1367-2630-15-7-073043}.
Let us consider the single-particle map
\be
\Lambda_\phtheta[\varrho]=\sum_k E_k(\phtheta) \varrho E_k(\phtheta)^\dagger,
\ee
where $E_k$ are the usual POVM elements. 
Let us assume that this map acts on the input state on all particles independently, and we use the state obtained from this map 
to estimate the parameter $\phtheta.$ Thus, while before we estimated the parameter in a unitary evolution, now we 
need to estimate the parameter of the incoherent dynamics. 
The corresponding quantum Fisher information will be 
denoted by $F_Q[\Lambda_{\phtheta}^{\otimes N}[\varrho]].$

According to \REF{escher2011general},  the quantum Fisher information for this incoherent evolution can be computed as  
\be\label{FQbound}
F_Q[\Lambda_{\phtheta}^{\otimes N}[\varrho]]=\min_{\ket{\Psi_\phtheta}}
F_Q[\ket{\Psi_{\phtheta}}].
\ee
On the right-hand side, there is a minimisation carried out over all the purifications of the dynamics $\ket{\Psi_{\phtheta}}$ for which
\be
\Lambda_{\phtheta}^{\otimes N}[\varrho]={\rm Tr_E}(\ketbra{\Psi_{\phtheta}}),
\ee
where ${\rm Tr_E}(.)$ means tracing out the environment and the extended system consists of the original system and the environment.
The minimisation in \EQ{FQbound} is analogous to the infimum in the convex roof construction of \EQ{convroof}.
If we find a purification which is not the optimal one, based on \EQ{FQbound}, it will still give an upper bound on the
quantum Fisher information, which can be used to prove the shot-noise scaling of particular
quantum metrological setups. 

\REFL{demkowicz2012elusive} presented an alternative proof based on
the classical simulation of channels. Let us assume that a single-particle channel is obtained as a mixture of other channels as
\be
\Lambda_{\phtheta}[\varrho]=\int {\rm d}x p_{\phtheta}(x) \Lambda_x[\varrho].
\ee
Then, the Cram\'er-Rao bound gives a lower bound
on the variance of the parameter to be estimated as
\be
(\Delta \phtheta_N)^2 \ge \frac{1}{NF_{\rm cl}[p_{\phtheta}]},\;\;\;
F_{\rm cl}=\int {\rm d}x \frac{[\partial_{\phtheta}p_{\phtheta}(x)]^2}{p_{\phtheta}(x)},
\ee
where $F_{\rm cl}$ is the classical Fisher information. Note that $F_{\rm cl}$
characterises the single-particle channel, thus it does not depend on $N.$
Thus, if $F_{\rm cl}<\infty$  then we obtain $(\Delta \phtheta_N)^2 \ge \mathcal{O}(\frac{1}{N})$ which means
a shot-noise scaling for large $N.$ It can be shown that $F_{\rm cl}<\infty$ 
for all channels that are not $\phtheta$-extremal. The behaviour of such channels for small $\phtheta$ can always be
described by the mixture of two channels that are at the boundary of the set of channels.
A third proof in \REF{demkowicz2012elusive}, working also for  $\phtheta$-extremal channels, makes it possible to avoid the optimisation over purifications by  carrying out the optimisation in an extended system.

\REFL{PhysRevLett.112.120405} presents a general formalism for  open system
metrology, which is able handle complicated 
non-unitary dynamics and provides a lower and an upper bound
for the quantum Fisher information. In general, there has been a large recent interest in parameter estimation in non-unitary processes, however, this topic is outside of the scope of the review (e.g., see \REFS{PhysRevLett.100.100503,PhysRevLett.98.160401}). 

Due the uncorrelated noise, the states obtained can efficiently be  described by matrix product states \cite{PhysRevLett.110.240405}.
The change of the scaling of the precision to a shot-noise scaling for large $N$ has been observed also in the 
analysis of the scaling for the squeezed-light-enhanced gravitational wave detector GEO 600 \cite{PhysRevA.88.041802}.

There have been several attempts to beat the shot-noise scaling in the large particle number limit 
even in the light of the above results. For example, 
in metrology with GHZ states described in \SEC{sec:Metrology with a GHZ state}, \REF{PhysRevLett.111.120401} considers a frequency measurement such that there is an optimisation over the duration of the dynamics. Hence, the single-particle channel depends on $N,$ while it was independent of $N$ in the discussion above.
It is shown that  the shot-noise limit can be surpassed for a particular type of dephasing, concretely, for the $\alpha_x=1$ case in the decoherence model given in \EQ{eq:LK}.
Note, however, that the shot-noise scaling is restored for large N whenever $\alpha_x<1,$ e.g., if $\alpha_z>0.$
Moreover, very recently, quantum error correction has been used in proposals to reach a Heisenberg scaling in a noisy environment \cite{PhysRevLett.112.150802,PhysRevLett.112.080801}. 

So far we have been discussing metrological setups for an ensemble of particles that do not interact with each other.
Such setups play an important role in metrology, as it is much easier to create dynamics with non-interacting particles in a controlled 
manner, than dynamics based on two-body interactions. 
With interactions, the precision can surpass 
the shot-limit of the non-interacting case even starting from a product state, or it can surpass the Heisenberg limit of the non-interacting case \cite{Luis20048,napolitano2011interaction,PhysRevLett.98.090401,braun2011heisenberg,PhysRevLett.100.220501,PhysRevA.77.053613,PhysRevA.76.053617}.
It is instructive to obtain the
 maximal scaling for systems with $q$-body interactions as follows.
Let us take the Hamiltonian
$H_q=J_x^q.$ For this Hamiltonian, 
$F_Q[\varrho,H_q]/4\le\va{H_q}\le(\frac{N}{2})^{2q}=\mathcal{O}(N^{2q}).$ 
Similar derivation works also for any Hamiltonian with at most $q$-particle interactions.
However, for a wide class of noise models,
uncorrelated noise affects these setups in way analogous to the linear case:
due to the uncorrelated noise, only $\mathcal{O}(N^{(2q-1)})$ can be reached for large $N$ \cite{PhysRevA.89.022107}. 

All the statements above are for uncorrelated noise. The case of correlated noise has also been studied intensively (e.g., see \REFS{dorner2012quantum,PhysRevA.84.012103,PhysRevLett.109.233601,knysh2014true}).  For certain types of correlated noise, the Heisenberg scaling can be reached.

Finally, it is important to recognise that our observations concerning the effect of the noise in quantum metrology are connected to fundamental questions in quantum physics. In \SEC{sec:Multipartite entanglement}, we have shown that for a large metrological precision a large entanglement depth is also needed. 
Naturally arises the question whether it is possible to reach a large entanglement depth, especially for large particle numbers in a noisy environment. Hence, quantum metrology is connected to several other fields examining the survival of large scale entanglement in a noisy  environment  \cite{PhysRevLett.110.240405}. 
For example, a related topic is the physics of nanosystems at a finite temperature  \cite{PhysRevLett.93.080402,PhysRevA.79.052340}. For such systems, the density matrix of a system is very close to a tensor product of density matrices of subsystems. The larger the temperature, the smaller the size of the terms can be. This is connected to the fact that multipartite entanglement cannot survive easily at finite temperatures. 

\section{Conclusions}

We have discussed the basics of quantum metrology through simple examples 
of metrology with a fully polarised ensemble of particles or with highly entangled states.
After this introduction, we presented the basic formalism of quantum metrology based
on the quantum Fisher information and the Cram\'er-Rao bound.
We discussed that for dynamics with a Hamiltonian that does not contain interaction terms, the usefulness
of quantum states for metrology is closely related to their entanglement properties.
We found that separable states can achieve only a shot-noise scaling, while
for states with high level of entanglement even the Heisenberg scaling is possible.
Finally, we discussed how uncorrelated noise can affect this situation, leading back to a shot-noise scaling for large particle numbers even
for entangled states. 

\section*{Acknowledgments}

We thank S.~Altenburg, E.~Bagan, N.~Behbood, J.~Calsamiglia, G.~Colangelo, M.~Cramer, R.~Demkowicz-Dobrza{\'n}ski, F.~Deuretzbacher, B.~Escher, I.~L.~Equsquiza, F.~Fr\"owis, O.~G\"uhne, P.~Hyllus, M.~Kleinmann, C.~Klempt, B.~Kraus, B.~L\"ucke, M.~W.~Mitchell, M.~Modugno, A.~Monras, M.~Oberthaler, D.~Petz, L.~Pezz\'e, L.~Santos, C.~Schwemmer, R.~J.~Sewell, A.~Smerzi, R.~Schmied, and H.~Weinfurter for interesting discussions. 
We thank O. G\"uhne for calling our attention to the entanglement of assistance measures for \SEC{Sec:Quantum Fisher information},
and P. Hyllus for discussions about entanglement in systems of very many particles connected to \SEC{Sec:Multi-particle entanglement vs. mode-entanglement}.
A special thank to the members of the group funded by the ERC StG GEDENTQOPT at the University of the Basque Country UPV/EHU, to I.~Urizar-Lanz, G.~Vitagliano, and Z.~Zimbor\'as
 for discussions. Finally, we would like to thank the participants
 of the conference on Entanglement Detection and Quantificaton held in Bilbao in 2014 for many
 valuable discussions.
 We acknowledge the support of
  the EU (ERC Starting Grant GEDENTQOPT, CHIST-ERA QUASAR)
the Spanish MINECO
(Project No. FIS2009-12773-C02-02 and No. FIS2012-36673-C03-03),
the Basque Government (Project No. IT4720-10), and
the National Research Fund of Hungary OTKA (Contract No. K83858).

\section*{References}

\bibliography{metrorev_links}

\end{document}